\documentclass[11pt,amsmath,amssymb,onecolumn]{article}

\usepackage{graphicx}
\usepackage{amsmath}
\usepackage{amsfonts}
\usepackage{xcolor}

\usepackage[toc,page]{appendix}

\usepackage{scicite}

\usepackage{times}

\newenvironment{sciabstract}{%
\begin{quote} \bf}
{\end{quote}}

\newcounter{lastnote}

\title{Inverse Design of All-dielectric Metasurfaces with Bound States in the Continuum} 
\author
{Sergei Gladyshev$^{1,\ast,\dagger}$, Theodosios D. Karamanos$^{2,\ast,\dagger}$, Lina Kuhn$^{3,4}$,\\ Dominik Beutel$^{3}$, Thomas Weiss$^{1}$, Carsten Rockstuhl$^{3,5}$, Andrey Bogdanov$^{6}$\\
\\
\small{$^{1}$Institute of Physics, University of Graz, Universitätsplatz 5, 8010 Graz, Austria,}\\
\small{$^{2}$Institut Langevin, ESPCI Paris, Université PSL, CNRS, 75005 Paris, France,}\\
\small{$^{3}$Institute of Theoretical Solid State Physics, Karlsruhe Institute of Technology, 76131 Karlsruhe, Germany,}\\
\small{$^{4}$Steinbuch Centre for Computing - Scientific Computing \& Mathematics, Karlsruhe Institute of Technology, Germany}\\
\small{$^{5}$Institute of Nanotechnology, Karlsruhe Institute of Technology, 76344 Eggenstein-Leopoldshafen, Germany}\\
\small{$^{6}$Qingdao Innovation and Development Base of Harbin Engineering University, Qingdao 266000, Shandong, China}\\
\\
\small{$^\ast$e-mail: sergei.gladyshev@uni-graz.at, theodosios.karamanos@espci.fr.}
\\
\small{$^\dagger$These authors have contributed equally.}
}
\date{}
\begin{document} 
%\baselineskip24pt
\baselineskip14pt
\maketitle 
\begin{sciabstract}
% Abstract - empty
Metasurfaces with bound states in the continuum (BICs) have proven to be a powerful platform for drastically enhancing light-matter interactions, improving biosensing, and precisely manipulating near- and far-fields. However, engineering metasurfaces to provide an on-demand spectral and angular position for a BIC remains a prime challenge. A conventional solution involves a fine adjustment of geometrical parameters, requiring multiple time-consuming calculations. In this work, to circumvent such tedious processes, we develop a physics-inspired, inverse design method on all-dielectric metasurfaces for an on-demand spectral and angular position of a BIC. Our suggested method predicts the core-shell particles that constitute the unit cell of the metasurface, while considering practical limitations on geometry and available materials. Our method is based on a smart combination of a semi-analytical solution, for predicting the required dipolar Mie coefficients of the meta-atom, and a machine learning algorithm, for finding a practical design of the meta-atom that provides these Mie coefficients. Although our approach is exemplified in designing a metasurface sustaining a BIC, it can, also, be applied to many more objective functions. With that, we pave the way toward a general framework for the inverse design of metasurfaces in specific and nanophotonic structures in general.
%
%Engineering metasurface with optical properties on-demand by techniques from the field of inverse design is an  important and challenging task in nanophotonics. In particular, identifying metasurfaces that offer bound states in the continuum (BIC) - resonances with extremely large quality factors - for a given illumination would have a tremendous impact on the future development of applications. While frequently considered as an emerging feature, we suggest in this contribution a deterministic approach. We exploit techniques from the field of machine learning to identify the geometry and material parameters of an all-dielectric metasurface so that it supports a parametric BIC at a predefined frequency and incidence angle of a plane wave used for illumination. Our method starts with a parametrization of the possible response of a scatterer that decorates the unit cell using Mie angles. While considering the full lattice interaction, a scatterer can be identified that offers the desired BIC. Then, an artificial neural network predicts the geometrical and material properties of a core-shell meta-atom that offers the necessary optical response to cause a BIC at the desired frequency and angle of incidence. While our approach is exemplified in the design of metasurface sustaining a BIC, it can be applied to many more objective functions. With that, we pave to way toward a general framework for the inverse design of metasurfaces in specific and nanophotonic structures in general.
\end{sciabstract}
\section*{INTRODUCTION}
Bound states in the continuum (BICs) are non-radiating solutions to the wave equation with a spectrum embedded in the continuum of the propagating modes in the surrounding space. BICs are a general wave phenomenon that can exist in a variety of acoustic, hydrodynamic, quantum mechanical, and electromagnetic systems~\cite{Hsu2016Jul,koshelev2019nonradiating,KoshelevKL2023Feb}. Because of their infinite radiative lifetimes, BICs are actively studied in optics and photonics, opening up enormous opportunities to realize compact planar high-Q metastructures necessary for biosensing, integrated nonlinear nanophotonics, and an enhancement of light-matter interactions~\cite{Azzam2021Jan}. In photonics, the most promising platform supporting BICs are {\it metasurfaces}~\cite{Qiu2021Jul}. Metasurfaces offer a vast number of designs with materials compatible to BIC. In particular, metasurfaces with BICs demonstrated their efficiency for lasing~\cite{Kodigala2017Jan,Ha2018Nov,Huang2020Feb,Hwang2021Jul,Azzam2021Mar}, biosensing~\cite{Ndao2020May,Tittl2018Jun,Leitis2019May,Romano2018Aug}, polaritons~\cite{Kravtsov2020Apr,Cao2020Jun,maggiolini2022strongly,weber2022strong}, and enhancing nonlinear optical effects~\cite{Bernhardt2020Jul,Zograf2022Feb,Sinev2021Oct,Liu2021Sep}.

Optical resonances in periodic metasurfaces radiate only into the open diffraction channels, while BICs remain non-radiating due to the vanishing coupling to all open diffraction channels. For subwavelength metasurfaces, there is only one open diffraction channel and, thus, one coupling coefficient.  It can vanish due to symmetry reasons or due to a fine-tuning of the system's geometrical or material parameters~\cite{Zhen2014Dec}. In the first case, the BICs are called {\it symmetry-protected} and usually exist in high symmetry points of the $k$-space. In the second case, BICs are called {\it accidental} or {\it parametric}, and they can exist at an arbitrary point in the $k$-space along the high-symmetry directions~\cite{Hsu2013Jul}. This explains the term ``accidental''. The accidental BICs are extremely sensitive to changes in the geometry of the unit cell or the material parameters. Such a high sensitivity makes it challenging and time-consuming to design metasurfaces with a pre-required or fixed spectral and angular position of BIC. 

%The spectrum of BICs is embedded in the radiation continuum of propagating modes in the surrounding space. However, BICs remain non-radiating due to the vanishing coupling to the scattering channels. The scattering channels can be identified in periodic metasurfaces with diffraction orders~\cite{Zhen2014Dec}. The coupling between the BICs and the diffraction orders can vanish due to symmetry reasons or due to a fine-tuning of the system's geometrical or material parameters. In the first case, the BICs are called {\it symmetry-protected} and usually exist in high symmetry points of the $k$-space. In the second case, BICs are called {\it accidental} or {\it parametric}, and they appear at an arbitrary point in the $k$-space along the high-symmetry directions~\cite{Hsu2013Jul}. This explains the term ``accidental''. Accidental BICs are extremely sensitive to changes in the geometry of the unit cell or the material parameters. Such a high sensitivity makes it challenging and time-consuming to design metasurfaces with a pre-required or fixed spectral and angular position of BIC. 

As an alternative to the time-consuming and brute-force  optimization~\cite{evlyukhin2021polarization,abujetas2022tailoring}, an {\it inverse design} approach can be used~\cite{Wiecha2021May,Molesky2018Nov,elizarov2022inverse,wiecha2022inverse}. The term ``inverse design'', herein, refers to the process of designing a metasurface with optical properties by initially specifying the desired response rather than the structure of the metasurface itself. Such an inverse design involves dedicated algorithms, e.g., from Bayesian inference, topology optimization, or artificial neural networks. The purpose is always to find the optimal metasurface that provides a predefined optical response~\cite{Jiang2021Aug,Ma2021Feb,Krasikov2022Mar,estrada2022inverse}. Inverse design can offer metasurfaces with a wide range of optical properties, including phase shifts, polarization conversions, and beam steering. It is a powerful tool for designing advanced optical devices and has applications in fields such as imaging, sensing, and telecommunications~\cite{Zahavy2018May,Kabir2008Apr,Zhang2019Mar,Rivenson2017Nov}. Despite the power of inverse design methods based on artificial neural networks, the underlying physics of the found optimum often remains vague. 

\begin{figure*}[ht!]
\begin{center}
\includegraphics[width=0.9\textwidth]{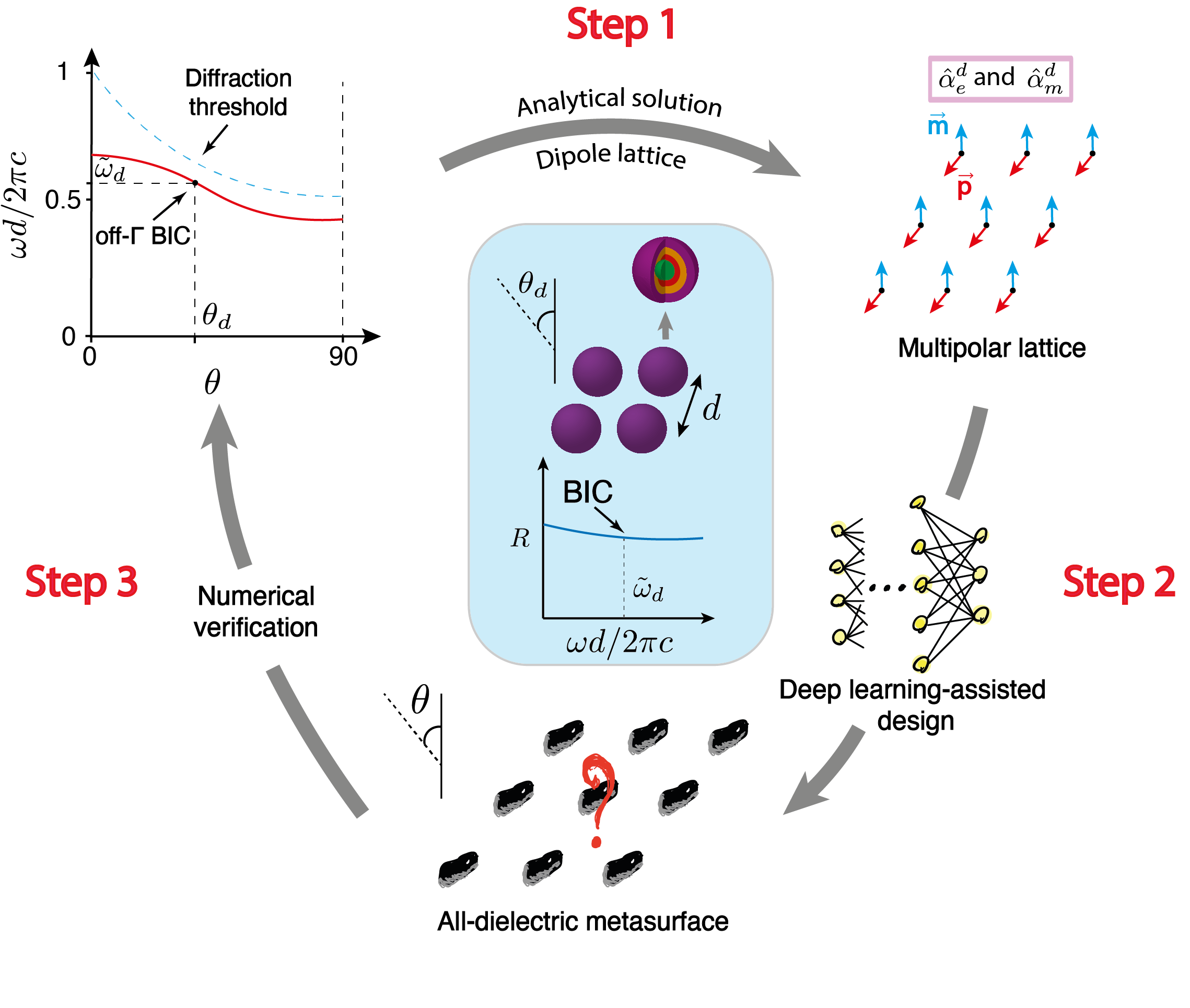}
\caption{The concept of our proposed framework. Beginning from the top left, an accidental BIC sustained by a metasurface is desired at a specific normalized frequency and incidence angle given by $(\tilde{\omega}_d, \theta_d)$. Then, the metasurface is analytically modeled via the multipolar expansion up to dipolar order. The predefined BIC is supported when the isotropic meta-atom decorating the unit cell of the metasurface has a specific electric and magnetic dipolar polarizability. Following that, a realistic core-shell particle that provides the desired polarizabilities is designed via a machine-learning algorithm in the second step. Finally, the proposed, realistic metasurface is numerically validated concerning its performance and robustness in the third step. By the latest at that stage, a realistic metasurface is described in all its details.} 
\label{fig:concept}
\end{center}
\end{figure*}

In this work, we develop a physics-inspired inverse design procedure for all-dielectric metasurfaces sustaining an accidental BIC at a predefined frequency and incidence angle. Our framework is based on a smart combination of a semi-analytical approach and a dedicated machine-learning algorithm. Our procedure includes three steps schematically shown in Fig.~\ref{fig:concept}. 
In the first step, the problem is solved within a ``toy'' model. Here, we consider all-dielectric meta-atoms in dipole approximation and capture their response using a T-matrix. After representing the meta-atoms utilizing the electric/magnetic (dipole) polarizabilities, or, equivalently, the scattering (dipole) Mie coefficients, the metasurface model is set up. The existence condition for an accidental BIC at a predefined spectral and angular position is derived analytically as a function of the Mie coefficients of the constituting particle through the identification of the system's eigenmodes. At the end of that first step, we know the polarizabilities of the particle such that the metasurface offers the predefined BIC. 
In the second step, using a dedicated artificial neural network, we identify the geometrical and material parameters of a core-shell, spherical particle that offers the desired and previously identified polarizabilities so that the metasurface sustains the predefined BICs. The neural network accounts for the practical limitations on the sizes of the spherical particle shells and refractive indices of the materials. In this way, a physically feasible meta-atom is found that provides the desired BIC when placed on an infinite 2D array. 
In the last step, we numerically verify the design. We also test the robustness of the designed BIC position in $k$-space if higher-order multipoles contribute besides the dipolar one, namely quadrupoles and octupoles. This entire process is described in the following. 
\section*{RESULTS}
\subsection*{Step 1: BIC in dipolar metasurface}

\begin{figure*}[ht!]
\begin{center}
\includegraphics[width=1\textwidth]{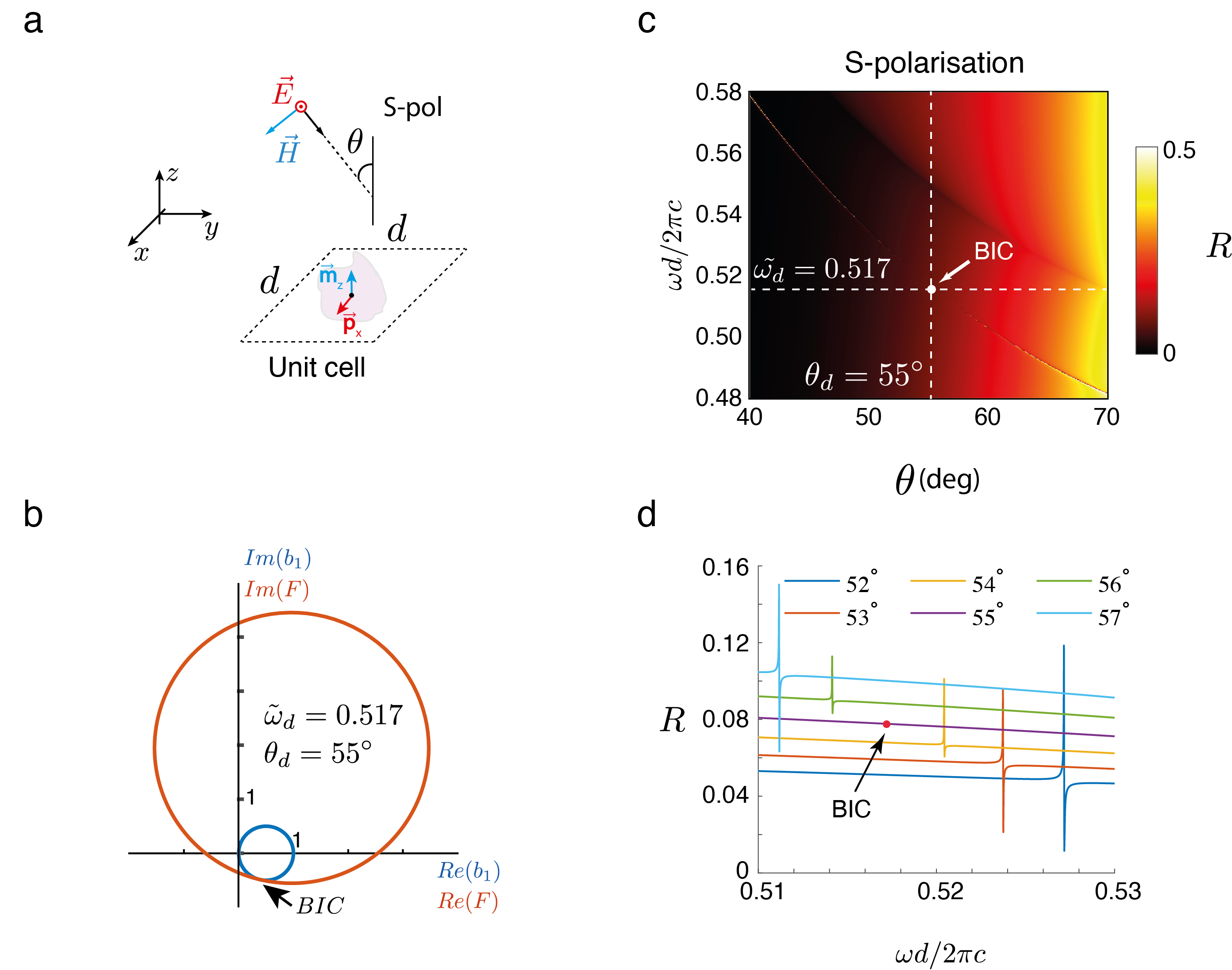}

\caption{The toy model: (a) A unit cell of a metasurface made from a periodic arrangement of scatterers with electric and magentic dipole polarizabilities  in a square lattice. (b) A graphical representation of how the Mie coefficients are identified that lead to the desired response. The analytical expression for the Mie coefficient that needs to be satisfied to sustain a BIC at a predefined normalized frequency and incidence angle for a given period is plotted separately concerning its left- and right-handed sides. Each side depends only on one Mie coefficient. By parametrizing the Mie coefficients with the Mie angles, both sides of the expression in the complex plane have been plotted. From the point of crossing, the Mie coefficients that provide the desired BIC are identified. (c) The reflection $R$ as a function of dimensionless frequency $\omega d /2 \pi c$ and angle of incidence $\theta$ for a square lattice with period $d = 450$ nm. The lattice is decorated with particles that offer the previously identified Mie coefficients. The appearance of the BIC at the predefined frequency and incidence angle can be seen. (d) The reflection $R$ as a function of dimensionless frequency $\omega d /2 \pi c$ in close proximity to the off-$\Gamma$ BIC. } 
\label{fig:toy}
\end{center}
\end{figure*}

Let us consider in our "toy" model a metasurface consisting of isotropic and non-absorbing particles described in dipole approximation. The particles are arranged on a square lattice with a subwavelength period, i.e. for any incidence angle, there is only a zeroth diffraction order. The considered unit cell is shown in Fig.~\ref{fig:toy}(a). The surrounding medium is a vacuum. The metasurface is illuminated by a time-harmonic linearly polarized plane wave in either TE (or s) or TM (or p) polarization.

Considering the renormalization of the particle's polarizability due to the lattice interaction and imposing a condition that expresses the existence of a resonance, i.e., a denominator in the renormalized polarizability has to be zero, allows for the analytical identification of an equation that must be satisfied for the existence of a BIC. For a given lattice, frequency, polarization, and incidence angle, this equation expresses the necessary relation between the dipolar magnetic $b_1$ and electric $a_1$ Mie coefficients to encounter a BIC. A derivation is given in the METHODS section, but the final expressions read
\begin{subequations}\label{bic-formulas-dipole-iso}
\begin{align}
&b_1 = -\frac{1 + a_1\left(C_{1}+C_{3}\right)}
{C_{2} + a_1\left(C_{1}C_{2}+C_{2}C_{3}-2\,C_{5}^{\,2}\right)}=F(a_1),\quad &\text{(TE / s-polarized incidence)} \label{bic-formulas-dipole-iso-a}\\
&a_1 = -\frac{1 + b_1\left(C_{1}+C_{3}\right)}
{C_{2} + b_1\left(C_{1}C_{2}+C_{2}C_{3}-2\,C_{5}^{\,2}\right)}, \quad &\text{(TM / p-polarized incidence)} \label{bic-formulas-dipole-iso-b}
\end{align}
\end{subequations}
%
%where $C_{1} = C^{\, vv}_{-1-1} = C^{\,vv}_{11}$, $C_{2} = C^{\,vv}_{00}$, $C_{3} = C^{\,vv}_{1-1} = C^{\,vv}_{-11}$ and $C_{5} = C^{\,vv'}_{10} = C^{\,vv'}_{01}$, with $\{v,v'\} = \{ {\rm e,m} \}$ and $v \neq v'$, are the dipole-dipole interaction coefficients 
where $C_{i}$ ($i=1,...,5$) are the elements of the lattice interaction coefficients matrix $\bar{\bar{C}}_s$ (see Supplementary Material, Sec.~S5).
These interaction coefficients are calculated via an Ewald summation methods~\cite{beutel2021efficient,rahimzadegan2022comprehensive,beutel2023unified}. 

The elements of $\bar{\bar{C}}_s$ depend on the lattice constant $d$, frequency $\omega$, and Bloch wavevector. Therefore, they carry the information about the desired point of operation. Once they are fixed, the Mie coefficients can be identified such that a BIC is supported.
Note that the equations for s- and p-polarization are similar. Only the coefficients $a_1$ and $b_1$ are swapped, as anticipated, due to symmetry. One should notice from \eqref{bic-formulas-dipole-iso} that the electric-magnetic lattice coupling coefficient is crucial for the existence of a simultaneous solution for $(a_1,b_1)$, indicating the importance of multipolar, electromagnetic coupling to realize BICs.
The analytical derivation of Eqs. \eqref{bic-formulas-dipole-iso} is further elaborated in the \textit{Supplementary Material}.

Hence, for the specific scenario considered, Eqs \eqref{bic-formulas-dipole-iso} provides the exact condition to encounter a BIC for s- or p-polarized incidences, respectively. To further simplify the design, the Mie coefficients are parametrized using what is called the Mie angles \cite{rahimzadegan2020minimalist}. For a lossless particle, a single angle bound between $-\pi/2$ and $\pi/2$ is sufficient to express any possible value a Mie coefficient might attain. A representation of the Mie coefficients in terms of these Mie angles is highly beneficial for the further design. For example, for the s-polarized incidence case, by substituting dipole electric and magnetic Mie angles, $\theta_{\rm E1}$ and $\theta_{\rm M1}$ (see METHODS) into Eq. \eqref{bic-formulas-dipole-iso-a},
the exact solution can be easily obtained via a non-linear equation solver for a given wavelength, incidence angle, and lattice dimension.
%Alternatively, one can sweep $\theta_{\rm E1}$ in expression \eqref{bic-formulas-dipole-iso-a} between $-\pi/2$ and $\pi/2$ and keep only the solutions that satisfy the condition $|b_1|=1$ for lossless particles\cite{rahimzadegan2020minimalist}, as performed in \cite{rahimzadegan2022comprehensive}.    

The process of identifying the BIC graphically can be demonstrated by assuming an s-polarized incidence, as depicted in Fig.~\ref{fig:toy}(a). After expressing the right hand side of Eq. \eqref{bic-formulas-dipole-iso-a} as ${\rm F}(a_1)$,All possible Mie angles in the range of $[-\pi/2, \pi/2]$ , which parametrize the magnetic and electric dipolar coefficients, are swept through. The left-hand side (only the $b_1$ coefficient) and the right-hand side of Eq. \eqref{bic-formulas-dipole-iso-a} are shown in the complex plane in Fig.~\ref{fig:toy}(b).
between $-\pi/2$ and $\pi/2$ that parametrize the magnetic and electric dipolar coefficients. The left-hand side, i.e. only the $b_1$ coefficient, and the right-hand side of \eqref{bic-formulas-dipole-iso-a} are shown in the complex plane in Fig.~\ref{fig:toy}(b).
As required by lossless scatterers, all possible Mie coefficients, including the ones of $b_1$, lie on a circle in the complex plane with a center at the $(0.5,0)$ point and a radius of $1$. It can be identified that the BIC is located where ${\rm Re}\{b_1\} = {\rm Re}\{{\rm F}(a_1)\}$ and ${\rm Im}\{b_1\} = {\rm Im}\{{\rm F}(a_1)\}$ for a specific wavelength, incidence angle, and lattice constant. Once the $b_1$ coefficient is known, the $a_1$ value can be explicitly calculated. 

As an example, we find a combination of Mie angles ($\theta_{E1}, \theta_{M1}$), or, in other words, $\bar{\bar{T}}_0$ matrix, for which a BIC exists for the desired normalized frequency $\tilde{\omega}_d = \omega\, d /2 \pi c = 0.517$ and incident angle $\theta_d = 55^{\circ}$ in TE- / s-polarization.
%It should be noted that this matrix $T_0$ has no frequency dependence. 
%The index d stands for "desired".
The reflection from a metasurface formed by the 2D array of dipolar particles, corresponding to the calculated $(a_1,b_1)$ values, is shown in Fig.~\ref{fig:toy}(c). The reflection is shown as a function of the normalized frequency, $\omega\, d /2 \pi c$, and the incidence angle, $\theta$. Additionally, we have been assuming here that the T-matrix is non-dispersive.

For the selected $a_1$ and $b_1$ dipole moments, a narrow resonant band is formed. This resonant band disappears precisely at the target parameters $(\tilde{\omega}_d,\theta_d)$, thus, hailing a BIC. To make the success of the proposed methodology more explicit, in Fig.~\ref{fig:toy}(d), we illustrate the change in resonance shape for the reflection coefficient, $R$, as a function of $\omega\, d /2 \pi c$, for different incident angles $\theta$ from $52^{\circ}$ to $57^{\circ}$. The resonance becomes infinitely narrow at an angle of $\theta_d$ at $\tilde{\omega}_d$, or the $Q$-factor becomes infinite. It implies that the incident wave does not interact with the metasurface that gets fully transparent. Therefore, the presented methodology succeeds in providing the Mie coefficients for a spherical particle to observe a BIC at a specific frequency and an angle of incidence of a plane wave upon a periodic arrangement of the particle.
\subsection*{Step 2: Deep learning-assisted engineering of spherical particles for BIC realization}
\begin{figure*}[ht!]
\begin{center}
\includegraphics[width=0.99\textwidth]{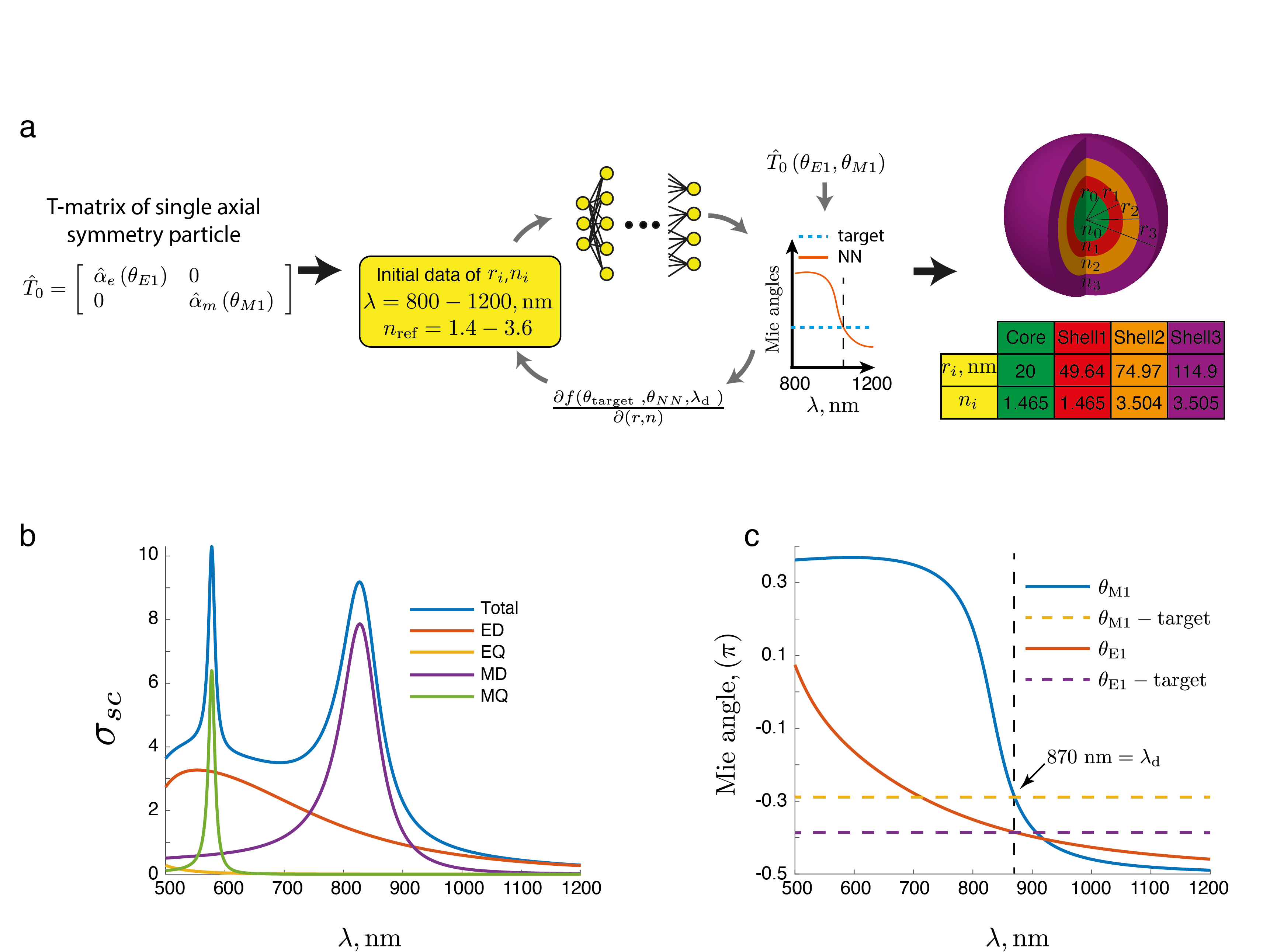}
\caption{Inverse design: (a) Scheme to find the geometrical and material parameters of a meta-atom (core-shell particle) with a target optical response. In the central part, there is a fully differentiable artificial neural network that can predict the Mie angles for a given core-shell particle. The network was trained within a given spectral region using discrete material classes and constrained geometrical dimensions for the core and the shell that make the design feasible for realization. A gradient descent is then used to identify the parameters characterizing the core-shell particle such that predefined dipolar Mie angles are provided at a design wavelength. At the very end, a second optimization is performed where the refractive indices of the considered materials are fine-tuned to reach an absolute precision. The table shows the design parameters for the example considered in the text. 
(b) Contribution of each multipole moment (up to quadrupolar order) to the scattering cross-section of the core-shell particle $\sigma_{sc}$ as a function of the wavelength $\lambda$ for the final design. (c)  Mie angles as a function of wavelength: the dotted line refers to the Mie angles of the optimized toy model, and the solid line refers to the Mie angles of the final design. The intersection of the dotted and solid lines in the $\lambda_d = 870\, \text{nm}$.} 
\label{fig:maps}
\end{center}
\end{figure*}
In the previous subsection, a method was presented that provides the T-matrix for an isotropic particle to achieve an accidental BIC for a specific normalized frequency and incidence angle.. Although this approach was successful, if one wants to provide a practical design of  a metasurface that exhibits BICs, the calculated $(a_1,b_1)$ values must be linked to a realistic scatterer. In this work, we employ a deep-learning scheme to assign the calculated Mie coefficients to a spherical core-shell nano-particle of realistic dimensions and made from existing materials for an operation at optical wavelengths.

To find a suitable scatterer that provides desired Mie angles, a gradient-based deep learning-assisted approach is performed, schematically explained in Fig.~\ref{fig:maps}(a). More details on the Artificial Neural Networks (ANNs) can be found in the article \cite{KuhnRepanRockstuhl2022}. Initially, we train a set of ANNs to predict the Mie angles of coated dielectric spheres made from one to five shells and at $200$ distinct wavelengths, between $800$~nm - $1200$~nm. A part of the ANNs is also a classifier that predicts the most probable number of shells to provide the requested Mie angles. Actually, the exemplarily shown core-shell particle in Fig.~\ref{fig:maps}(a) has only three shells. Our trained ANN simply predicted at the end a core-shell particle with three shells only as the most appropriate, which is illustrated here. The materials of the layers are restricted to discrete refractive indices classes ranging from $1.4 - 3.5$, values that are realistic at optical wavelengths for specific materials (see table~\ref{tab:materials}). %The first optimisation is discrete, second is continuous. This is to ensure that in the provided designs the position of the BICs match the desired ones. 
The radius of the core is limited to $20$~nm - $50$~nm. The dimension of each shell is restricted to $20$~nm - $40$~nm. The ANNs are used as a fully differentiable surrogate model in a gradient-based optimization algorithm, namely the limited-memory Broyden-Fletcher-Goldfarb-Shanno algorithm including boundary constraints, or L-BFGS-B \cite{byrd1995limited}.

\begin{table}[]
\begin{tabular}{|l|l|l|l|l|l|l|l|}
\hline
Class            & 1      & 2      & 3      & 4      & 5      & 6    & 7    \\ \hline
Refractive index & 1.4649 & 1.7196 & 1.9447 & 2.0745 & 2.4317 & 3.0    & 3.5  \\ \hline
Material        & SiO2   & MgO    & ZnO    & ZrO2   & TiO2   & AlAs & GaAs \\ \hline
\end{tabular}
\caption{Diferent materials, associated classes and refractive indices.}
\label{tab:materials}
\end{table}

The actual solution to the inverse problem starts from random particle parameters as network input, for which we predict the spectral dependencies of the Mie angles with the trained ANN. From that output, we compute an objective function $f$, the Mean Absolute Error (MAE) of the ANN output and the target Mie angles at the desired wavelength, $\lambda_{\mathrm{d}}$, as
\begin{eqnarray}
f(\theta_{\mathrm{target}}, \theta_{\mathrm{NN}}; \lambda_{\mathrm{d}}) & = & \frac{1}{2}\sum_{i\in\{{\rm E1,M1}\}}\left|\theta_{i, \mathrm{target}}(\lambda_{\mathrm{d}})-\theta_{i,\mathrm{NN}}(\lambda_{\mathrm{d}})\right|.
\end{eqnarray}
Subsequently, the gradients of $f$ with respect to the particle parameters $\frac{\partial f}{\partial (r,n)}$ are computed, and they are adjusted iteratively to minimize the MAE. This procedure is repeated for several initial parameters until we find a core-shell particle that provides the target Mie angles with high accuracy. 

Unfortunately, the discretization of the refractive index values naturally leads to a restriction of the possible design space. Thus, the design Mie angles and those of the optimized stricture do not match in a perfect sense after that procedure, i.e., we found an agreement only up to two digits after the comma. However, the BIC is very sensitive to slight changes in the Mie angles. Hence, we perform a second optimization using the actual analytical computation of the Mie angles and varied the refractive index of the involved materials slightly for fine-tuning. This approach alone is significantly slower than ANN-assisted design, especially for several initial trials. Fortunately, in this work, the designed particle can be fine-tuned to achieve the required accuracy (eight digits after the comma) and grant the appearance of the BIC, given the results of the first ANN-assisted optimization approach as a single starting point. The fine-tuning of the refractive index of the shells could be carefully done by suitable doping of the respective material, which should be within reach with existing technology~\cite{smietana2013capability,medda2005synthesis,schwenzer2012tuning}. 
In summary, this procedure provides a physical particle that can be used to form a metasurface that offers the BIC at the predefined frequency and incidence angle.
\subsection*{Step 3: Application of the design methodology}
The procedure presented above will now be applied to propose a metasurface with a square unit cell decorated by core-shell dielectric spheres that realize a BIC for a predefined lattice, frequency, and incidence angle. For this purpose, let us begin with the theoretical setup depicted in Fig.~\ref{fig:toy} with the resulting Mie coefficients from \eqref{bic-formulas-dipole-iso}, $a_1 = -0.4046 + 0.4908{\rm i}$ and $b_1 = -0.5508 + 0.4974{\rm i}$ for $\tilde{\omega}_d = 0.517$ and $\theta_d = 55^\circ$. Afterward, the algorithm presented in the previous section will be utilized to design a realistic core-shell spherical particle that provides the calculated Mie coefficients $a_1$ and $b_1$. We define the dimension of the square lattice as $d = 450$~nm, thus, the desired operational wavelength is $\lambda_d = 870$~nm. The obtained set of parameters of the core-shell particle, i.e., the refractive indices and the geometric dimensions of the shells and core, of which the square array should consist, is shown in Fig.~\ref{fig:maps}(a). 
Next, we perform a multipolar decomposition of the scattered field \cite{santiago2019decomposition} from the designed spherical particle upon illumination with a linearly polarized plane wave. As depicted in Fig.~\ref{fig:maps}(b), it possesses predominantly a dipolar response around the operational wavelength, $\lambda_d$. 
%Next, we perform a multipolar decomposition of the scattered field \cite{santiago2019decomposition} from the particle upon illumination with a linearly polarized plane wave and, as depicted in Fig.~\ref{fig:maps}(b), the designed spherical particle possesses predominantly a dipolar response around the operational wavelength, $\lambda_d$. 
Moreover, the Mie angles of the designed core-shell particle are calculated \cite{rahimzadegan2020minimalist} and presented for the wavelength spectrum $500 - 1200$~nm in Fig.~\ref{fig:maps}(c). Although the Mie angles are dispersive within the considered spectrum, they possess at the $\tilde{\omega}_d$ the required values for the designed operation, equal those requested by the theoretical model.
\begin{figure*}[ht!]
\begin{center}
\includegraphics[width=1\textwidth]{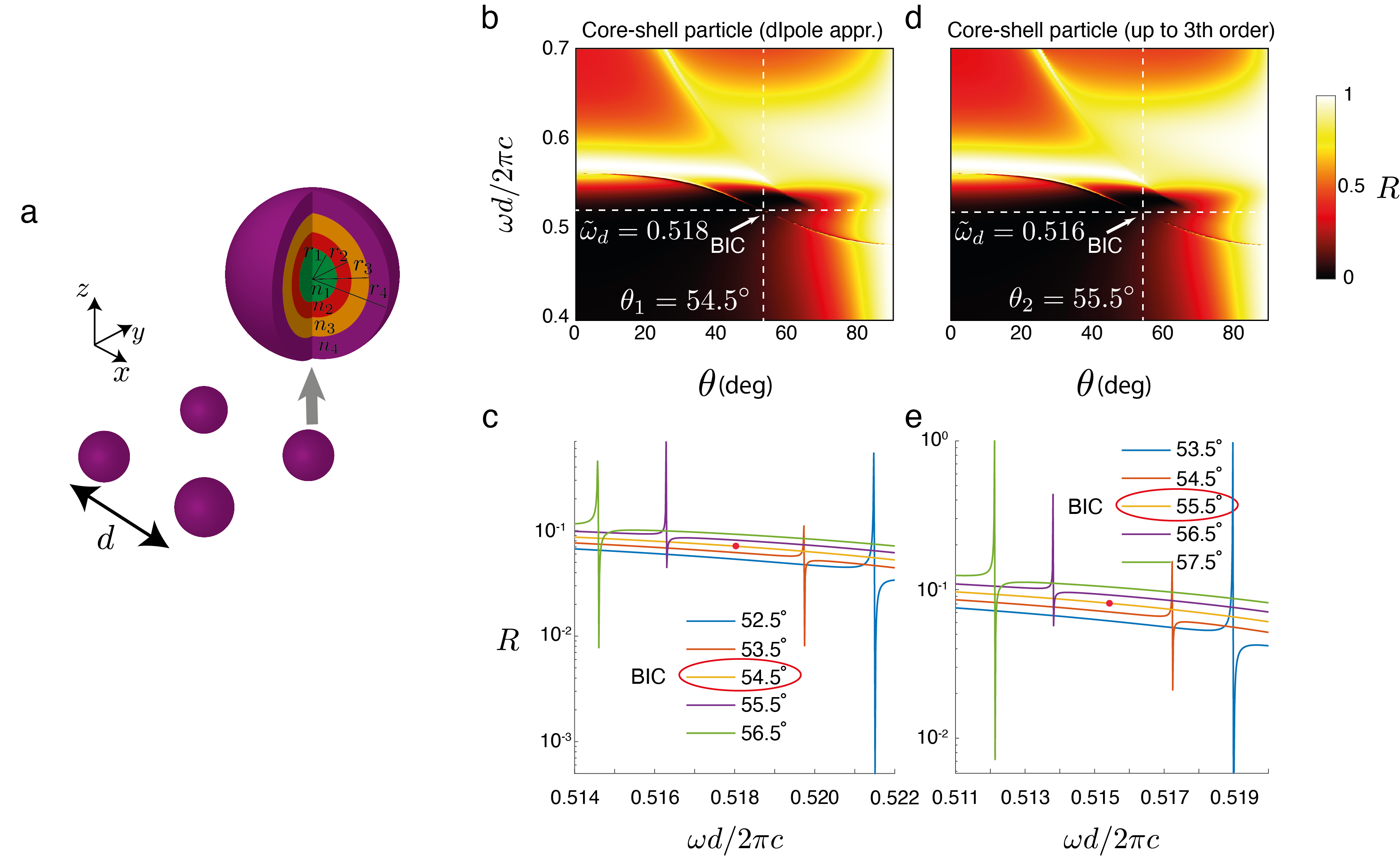}
\caption{(a) The square lattice of core-shell particles. (b,d) Reflection coefficients of the actual metasurface as calculated with a full-wave Maxwell solver that exploits the T-matrix formalism. The reflection is shown as a function of the frequency and the incidence angle for a lattice dimension $d = 450$~nm. (b) Calculation in dipole approximation. (d) Calculation in octupole approximation. (c,e) The reflection coefficient versus the normalized frequency in proximity to the off-$\Gamma$ BIC for a lattice dimension $d$ and using the dipole or the octupole approximation, respectively.} 
\label{fig:coreshell}
\end{center}
\end{figure*}

Finally, the designed core-shell spherical particle is placed on a 2D square array (Fig.~\ref{fig:coreshell}(a))
and the optical response from the metasurface is analyzed with a dedicated T-matrix based full-wave solver (See \textit{Supplementary Material}) \cite{beutel2021efficient,rahimzadegan2022comprehensive}.
In Figs.~\ref{fig:coreshell}(b) and (d), the reflection depending on the normalized frequency, $\omega d /2 \pi c$, and the incidence angle, $\theta$, are shown when only dipole (first order) and up to octupole (third order) multipoles are considered in the response calculation, respectively \cite{rahimzadegan2022comprehensive}. Please note, the dipolar approximation would correspond to the assumption in the design process. However, the actual particle does not, of course, have a purely dipolar response, but also small, yet non-negligible, higher-order multipolar coefficients. These higher-order multipolar contributions usually need to be considered when the response from an actual metasurface is predicted.  

Specifically, when only the dipolar response of the particle is considered (Fig.~\ref{fig:coreshell}(b)), a BIC is observed at the $(\tilde{\omega},\theta) = (0.518,54.5^\circ)$, almost exactly on the desired location in the parameter space. A more clear view of the reported BIC point is given in Fig.~\ref{fig:coreshell}(c), where the reflection coefficient of the corresponding metasurface is plotted versus the normalized frequency for various angles of incidence. Thus, the deviation of the obtained core-shell particle metasurface from the desired BIC point in the theoretical dipole analysis is $1 \%$. The discrepancy can easily be explained by the fact that the realistic structure does not provide exactly the desired Mie coefficients. In other words, the consideration of actual materials causes these small deviations. Finally, to verify that the proposed design still fulfills the set goals when higher order multipoles are taken into account, the reflection coefficient from the core-shell particle metasurface is calculated when considering multipoles up to octupolar order, i.e., $j=3$ \cite{beutel2021efficient,rahimzadegan2022comprehensive}. One can observe in Fig.~\ref{fig:coreshell}(d) and (e) that the resulting BIC position at $(\tilde{\omega},\theta) = (0.516,55.5^\circ)$ does not differ much from the desired one. Therefore, it can be safely deduced that the performance of the designed metasurface will remain the same in realistic conditions.  
\section*{DISCUSSION}
\begin{figure*}[t]
\begin{center}
\includegraphics[width=1\textwidth]{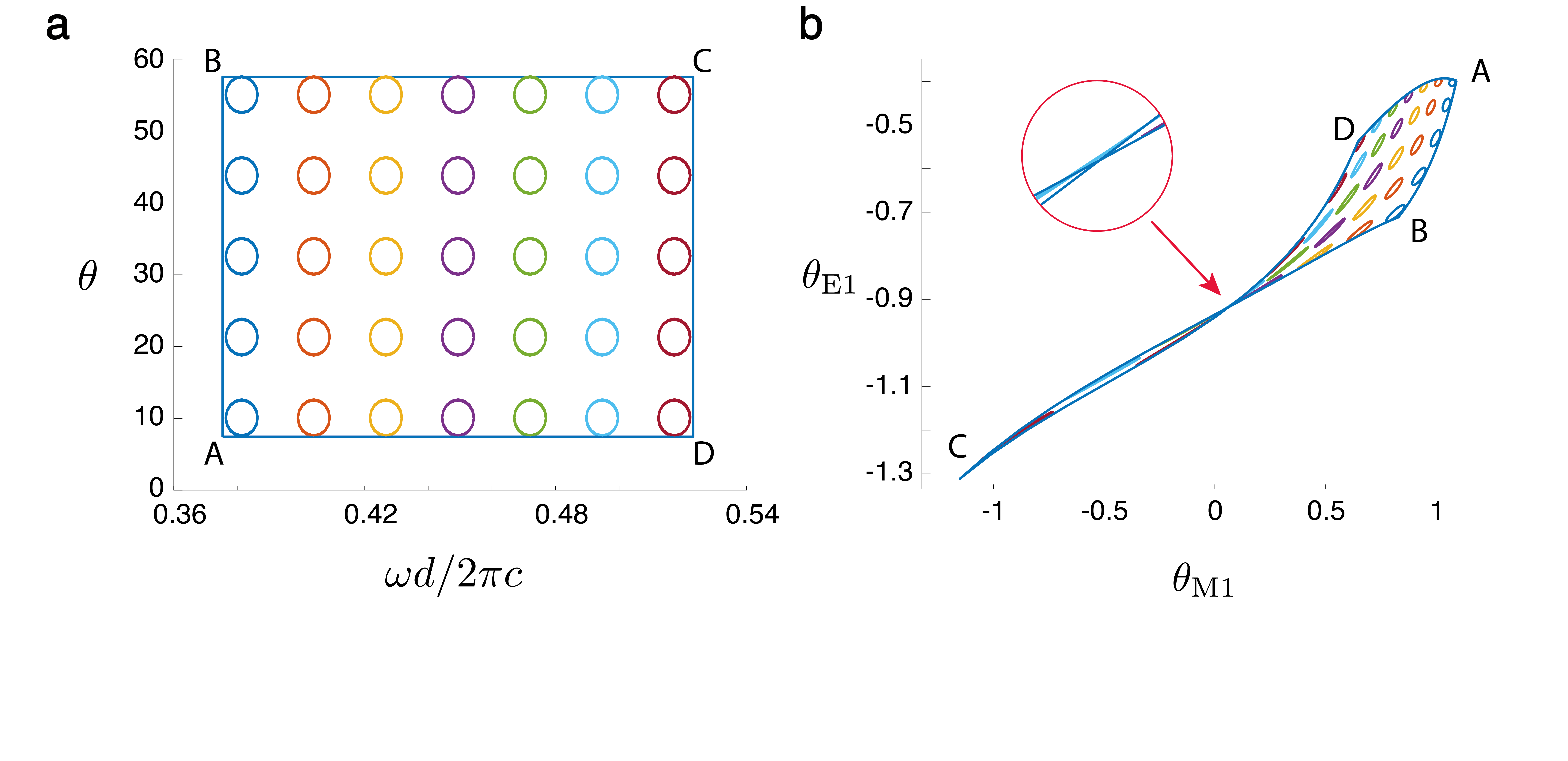}
\caption{Robustness of the BICs. (a) the positions of the BICs for different angles $\theta$ (deg.) and normalized frequency $\tilde{\omega}$ values: circles and contour in the form of a rectangle $ABCD$. (b) the values of the Mie angles for BICs from panel a.  } 
\label{fig:robustness}
\end{center}
\end{figure*}

The sensitivity of the BIC against deviations in the Mie angles parameterizing the optical response of the scatterers will now be discussed. Figure~\ref{fig:robustness} demonstrates how the position of the BIC for different angles $\theta$ and frequency $\tilde{\omega}$ values translates to values of the required Mie angles. The bounding box as a rectangle $ABCD$ is transformed very strongly in the plane of the Mie angles. That can also be seen in the individual circles that provide some insights into the robustness.
Generally, for a robust BIC, we require each area in the plane of the Mie angles to be as large as possible and as circular as possible. That would suggest that a given Mie coefficient only needs to be roughly hit to ensure that a BIC appears in the target area given by the circles in the plane spanned by the frequency and incident angle. However, the BIC is extremely sensitive for a strongly reduced area of the surface in the plane of the Mie angles. Thus, for that case, the Mie angles must be precisely met to cause a BIC at predefined frequencies and angles. 
Moreover, we notice that some circles are stretched and elongated in the plane of the Mie angles. Here, we will speak about conditional robustness. In this case, many Mie angles, for, let's say, the electric dipole moment, permit the observation of the BIC in the target range of parameters. However, the other Mie angle must be carefully tuned to be within the stretched circle. 
One can roughly conclude that the most robust region is in the zone of lower frequencies and shorter angles (close to the corner $A$). However, the excitations of BICs in the related parameter regime are difficult for practical systems. The sizes of realistic spheres, obtained as a result of the deep learning-assisted engineering step, compare already very well to the period of the metasurface. Therefore, the BIC position could shift a lot relative to the desired angles and frequencies since the renormalization is substantial in that parameter domain. 

In Fig.~\ref{fig:robustness}(b), the intersection of two lines of the bounding box occurs as a further interesting phenomenon. It means that for these Mie angles at the point of intersection, the metasurface sustains two distinct BICs for different pairs of frequency and incidence angles. This case is discussed in more detail in the \textit{Supplementary Material}. 

Although, in this work, we have utilized the dipole approximation to design the meta-atoms, i.e., only up to the first multipolar order, the core idea behind the presented methodology is general and can be applied to higher-order multipoles or more complex scatterers. Provided that the general eigenvalue problem (see \eqref{eig-problem-sph} in METHODS) can be solved under some approximations and the solutions corresponding to the BIC resonances can be recognized, the Mie angles that lead to the desired BIC can be expressed. For example, in \cite{rahimzadegan2022comprehensive}, for a square metasurface illuminated by a normally-incident plane wave and consisting of an isotropic particle with only a magnetic dipole and an electric quadrupole excited, a BIC point was identified, or the required $(b_1,a_2)$ values. Then, a suitable inverse design technique can be used to identify an actual meta-atom that offers these desired Mie angles. In the work at hand, the acquired Mie coefficients were assigned to a core-shell particle via the deep learning algorithm. These considerations can be extended to more complex, e.g., anisotropic or bi-anisotropic scatterers and can accommodate an increasing number of multipolar orders. Approximate models for the T matrix could be utilized for that purpose \cite{serdyukov2001electromagnetics,marques2011metamaterials} or more advanced neural networks can be trained \cite{repan2022exploiting}.

In summary, the physics-inspired inverse design procedure for all-dielectric metasurfaces sustaining an accidental BIC at a certain frequency and angle of incidence was developed in this work. The methodology utilized a smart combination of a semi-analytical approach and a dedicated machine-learning algorithm. Firstly, the problem was solved within a theoretical model considering all-dielectric meta-atoms using the dipole approximation and the T-matrix technique. The existence condition for an accidental BIC was derived in terms of Mie angles. Secondly, using a dedicated artificial neural network, the geometrical and material parameters of a core-shell, spherical particle that offered the desired optical response provided by the BIC condition were identified. In this way, a physically feasible meta-atom was found that provided the desired BIC when placed on an infinite 2D array. Finally, the design was numerically verified, and the robustness of the designed position BIC in $k$-space was tested by considering higher-order multipoles.
\section*{METHODS}
\subsection*{The T-matrix formulation for 2D arrays and BIC identification}
Considering a metasurface composed of a square array of isotropic and non-absorbing particles, as shown in Figure~\ref{fig:toy}(a). The surrounding medium is a vacuum. 

If an arbitrary scatterer is placed in an infinite homogeneous background and the vector spherical harmonics (VSH) basis is employed to expand the fields, the scattering response to an incident electromagnetic wave can be described using the T matrix, or $\bar{\bar{T}}_0$, as
\begin{equation} \label{t-matrix-def}
\left[\hspace{-0mm}\begin{array}{c}
\mathbf{a}^{\rm e}\\[0.05cm]
\mathbf{a}^{\rm m}
\end{array}\hspace{-0mm}\right] \hspace{-0mm} = \hspace{-0mm}\bar{\bar{T}}_0\hspace{-0mm}
\left[\hspace{-0mm}\begin{array}{c}
\mathbf{q}^{\rm e}\\[0.05cm]
\mathbf{q}^{\rm m}
\end{array}\hspace{-0mm}\right] = \left[\hspace{-0mm}
\arraycolsep=1mm
\begin{array}{cc}
\bar{\bar{T}}^{\,\rm ee} & \bar{\bar{T}}^{\,\rm em}  \\[0.05cm]
\bar{\bar{T}}^{\,\rm me} & \bar{\bar{T}}^{\,\rm mm}
\end{array}\hspace{-0mm}\right] \left[\hspace{-0mm}\begin{array}{c}
\mathbf{q}^{\rm e}\\[0.05cm]
\mathbf{q}^{\rm m}
\end{array}\hspace{-0mm}\right].
\end{equation}
Vectors $\mathbf{q}^{\{\rm e,m\}}$ contain the incident wave electric or magnetic coefficients, respectively, while $\mathbf{a}^{\{\rm e,m\}}$ contain the scattering electric or magnetic coefficients, respectively \cite{mishchenko2002scattering}. For each multipolar expansion order $j \in \mathbb{N}$, the size of the aforementioned vectors increases by $2j+1$.
Prior knowledge of the T matrix can predict the electromagnetic response of a scatterer, since it depends on the scatterer's geometry, constituting materials, and surrounding material. The elements of the T matrix can generally be extracted up to a preset expansion order via numerical simulations \cite{demesy2018scattering,santiago2019decomposition}, while for special geometries, such as spheres, cylinders, etc., the T matrix elements can be calculated analytically. 
The particular case of isotropic particles, i.e., spherical ones, is very interesting because apart from enabling analytic calculations, the respective T matrices become diagonal, or, $\bar{\bar{T}}^{\,\rm em} =  \bar{\bar{T}}^{\,\rm me} =  \bar{\bar{0}}$ and $T_{il}^{\,\rm ee} =  T_{il}^{\,\rm mm} = 0$, for $i\neq l$, with
$\{i,l\} \in [-j,j]$,
while $\bar{\bar{T}}^{\,\rm ee}_{j}$ and $\bar{\bar{T}}^{\,\rm mm}_{j}$ become scalar matrices for each multipolar order $j$, or, $\bar{\bar{T}}^{\,\rm ee}_{j} = -a_j\bar{\bar{I}}$ and $\bar{\bar{T}}^{\,\rm mm}_{j} = -b_j\bar{\bar{I}}$. The parameters $a_j$ and $b_j$ are called the \textit{Mie coefficients} \cite{Bohren2008}. %These properties of isotropic particles greatly simplify algebra and enable the exploration of physical phenomena, without loss of generality. 

For an isotropic particle, it is convenient to parameterize the coefficients with \textit{Mie angles} \cite{rahimzadegan2020minimalist}.
In particular, for lossless particles and when only dipoles are considered, i.e., $j=1$, all possible Mie coefficients can be expressed using two parameters $\theta_{{\rm E}1}$ and $\theta_{{\rm M}1}$, or
\begin{subequations}\label{mie-angles-def}
\begin{gather}
a_1 = \frac{1}{1-{\rm i}\,{\rm tan}\theta_{{\rm E}1}} , \,\, -\frac{\pi}{2}\leq\theta_{{\rm E}1}\leq \frac{\pi}{2},\\
b_1 = \frac{1}{1-{\rm i}\,{\rm tan}\theta_{{\rm M}1}} , \,\, -\frac{\pi}{2}\leq\theta_{{\rm M}1}\leq \frac{\pi}{2}.
\end{gather}
\end{subequations}
As a result, the optical response of the corresponding scatterer can be parameterized with 
the two Mie angles above, while still having a clear physical picture. This formulation is very important because the control parameters, namely $\theta_{{\rm E}1}$ and $\theta_{{\rm M}1}$, are placed between an upper and lower boundary, which will aid optimization methods and subsequent metasurface design. Furthermore, similar expressions involving Mie angles, like the ones of \eqref{mie-angles-def} above, exist for higher-order multipoles, the addition of absorption or Lorentzian dispersion for Mie coefficients (see \textit{Supplementary Material}). 
%or material dispersion. 
%Nevertheless, these formulations will not be employed herein and are presented in the \textit{Supplementary Material}, along with the expressions for particles with absorption, for any reader to expand our presented work.

Let us now assume an infinite square array composed of arbitrary, identical, absorption-less, spherical particles placed in a homogeneous material, as depicted in Fig.~\ref{fig:concept}. The response of the metasurface can be calculated via the T matrix of the particle by employing microscopic models \cite{xu2013scattering,antonakakis2014gratings,beutel2021efficient,rahimzadegan2022comprehensive}. An important aspect to be noted herein is that the T matrix of the particle within the metasurface is \textit{re-normalized} due to the particle interaction. Thus, in 2D array models, the isolated particle's T matrix, $\bar{\bar{T}}_0$, from \eqref{t-matrix-def} is replaced with the \textit{effective T matrix} calculated via \cite{xu2013scattering}
\begin{equation}\label{T-eff-def}
\bar{\bar{T}}_{\rm eff} = \left(\bar{\bar{I}} - \bar{\bar{T}}_0\bar{\bar{C}}_s\right)^{-1}\bar{\bar{T}}_0,
\end{equation}
where $\bar{\bar{I}}$ is the identity matrix, and $\bar{\bar{C}}_s$ is the \textit{lattice coupling matrix} expressed in spherical coordinates, which is a function of the unit cell dimension, the frequency, and the wavevector, and is calculated, herein, via rapidly converging summations using Ewald's method \cite{beutel2021efficient,beutel2023unified,fradkin2019fourier}. The analytical method based on a multipolar expansion that is used in this work for the calculation of the response of an infinite 2D square array \cite{rahimzadegan2022comprehensive} is elaborated in the \textit{Supplementary Material}.

It should be stressed that, although the scattering response of a single particle and of the subsequent 2D array is provided here using the spherical coordinates and the T matrix approach, the analysis that will follow is valid and interchangeable with the use of multipoles in Cartesian coordinates and the polarizability matrix. In particular, one can obtain the T matrix of a particle from the polarizability matrix, and vice versa using the appropriate transformation matrices for each multipolar order \cite{mun2020describing,rahimzadegan2022comprehensive}. Additionally, modifications of the same matrices can transform multipoles of the same type and order from a spherical to a Cartesian basis and vice versa, as presented in the \textit{Supplementary Materials} for multipoles up to the $j = 3$ order.

After defining the multipolar description of the electromagnetic response of 2D arrays composed of isotropic particles, next, we will present the procedure for identifying the presence of BICs after setting certain goals.
First, let us theoretically acquire the BIC position using the multipolar expansion technique  \cite{koshelev2020dielectric,abujetas2020coupled,evlyukhin2020bianisotropy,rahimzadegan2022comprehensive}. If we again assume a square array composed of identical and isotropic particles, its response to an incident field can be described by substituting \eqref{t-matrix-def} and \eqref{T-eff-def}  into the equations that describe the response of a 2D array \cite{beutel2021efficient,rahimzadegan2022comprehensive} (see \textit{Supplementary Material}), and, thus, a linear system of equations is formed. Solving the eigenvalue problem leads to the modes of the array, including, in this case, the trapped ones that do not couple with radiation channels, i.e., the BICs. Therefore, if we invert the square matrix of the system to the left side of \eqref{t-matrix-def} and set the excitation to zero, or $\mathbf{q}^{\rm\{ e,m\}} = \mathbf{0}$, the resulting homogeneous system will have a non-trivial solution if the determinant of the matrix is zero \cite{rahimzadegan2022comprehensive,abujetas2022tailoring,evlyukhin2021polarization}. In particular, after some algebra, the BIC condition is reduced to
\begin{equation}\label{eig-problem-sph}
\left|\,\bar{\bar{I}} - \bar{\bar{T}}_0\,\bar{\bar{C}}_s\,\right| = 0.
\end{equation}

The symmetry of the resonators is tightly related to their mode structure and multipole content, which determine the linear and non-linear response of the resonator. Using group theory, it is possible to classify the eigenmodes into irreducible representations and understand their multipole content \cite{sadrieva2019multipolar,gladyshev2020symmetry,xiong2020constraints}. Our model is characterized only by the dipole response.  Table 1 shows how the dipole moments correspond to different types of eigenmodes for a square array of particles outside the off-$\Gamma$ point in a spherical basis. The system is characterized by the $C_{2v}$ symmetry group in direction $\Gamma X$. There are only 4 types of modes: $A_1$, $A_2$, $B_1$, and $B_2$. Destructive interference of the eigenmodes of one type is necessary for BICs formed by the Friedrich-Wintgen mechanism. It can be concluded that in this model, parametric BIC exists only in modes of types $B_1$ and $B_2$. 
\begin{table}[t]
\begin{tabular}{|l|l|l|l|l|}
\hline
Irreducible representation & $A_1$     & $A_2$      & $B_1$                & $B_2$                  \\ \hline
Cartesian basis            & $p_y$      & $m_y$       & $p_z, m_x$           & $m_z, p_x$             \\ \hline
Spherical basis            & $a^\text{e}_{11}$, $a^\text{e}_{1-1}$ & $a^\text{m}_{11}$, $a^\text{m}_{1-1}$ & $a^\text{e}_{10}$, $a^\text{m}_{11}$, $a^\text{m}_{1-1}$ & $a^\text{m}_{10}$, $a^\text{e}_{11}$, $a^\text{e}_{1-1}$ \\ \hline
\end{tabular}
\caption{The classification of modes by irreducible representations and the multipole composition of the eigenmode for  $C_{2v}$ symmetry group in direction $\Gamma X$ in the dipole approximation for different bases.   }
\end{table}

The expression \eqref{eig-problem-sph} is general in nature and can be used for any type of lattice or particle in a homogenous medium \cite{rahimzadegan2022comprehensive}. Due to its complexity, \eqref{eig-problem-sph} can only be solved numerically in its general form, i.e., for higher-order multipoles or more diverse lattices. Nevertheless, for specific reduced cases, versatile analytic solutions can be found, as demonstrated in \cite{rahimzadegan2022comprehensive} for the case of the coupled electric dipole - magnetic quadrupole on a square lattice and a normal, TM wave incidence. In this work, we consider a square lattice decorated by isotropic and lossless particles, whose response is expanded only up to dipolar order, or $j=1$. Then, because the elements of the lattice interaction matrix, $\bar{\bar{C}}_s$, can be pre-calculated for a specific incident wavevector, $\mathbf{k}^{\rm inc}$, and a normalized lattice dimension, $d/\lambda$, eventually, \eqref{eig-problem-sph} can be solved for $j=1$ with the Mie angles of \eqref{mie-angles-def} as the unknowns. 
\section*{Acknowledgments}
	D.B. and C.R. acknowledge support by the Deutsche Forschungsgemeinschaft (DFG, German Research Foundation) under Germany’s Excellence Strategy via the Excellence Cluster 3D Matter Made to Order (EXC-2082/1-390761711) and by the Carl Zeiss Foundation. C.R. also acknowledges support by the Helmholtz Association via the Helmholtz program “Materials Systems Engineering” (MSE), and the KIT through the “Virtual Materials Design” (VIRTMAT). 
    L.K. acknowledges support by the NHR@KIT program and the Karlsruhe School of Optics and Photonics.

\section*{Supplementary material}
\textbf{Sections}\vspace{-1mm}
{\begin{itemize}
\item [-] S1. Field expansion via vector spherical harmonics. \vspace{-2mm}
\item [-] S2. Scattering from 2D arrays via multipolar expansion.\vspace{-2mm}
\item [-] S3. Mie angles definition.\vspace{-2mm}
\item [-] S4. Transformation from Cartesian to Spherical multipoles.\vspace{-2mm}
\item [-] S5. BIC identification for a dipole metasurface.\vspace{-2mm}
\item [-] S6. Computation of the lattice coupling matrix $\bar{\bar{C}}_s$. \vspace{-2mm}
\item [-] S7. Two BICs in one band.
\end{itemize}}

\noindent\textbf{Figures} \vspace{-1mm}
\begin{itemize}
\item [-] Fig.S1. A square lattice of identical particles along with the set-up Cartesian and spherical coordinate systems.\vspace{-2mm}
\item [-] Fig.S2. Model with a Lorentzian dispersion.\vspace{-2mm}
\item [-] Fig.S3. Two BICs on one mode line.\vspace{-2mm}
\end{itemize}

\noindent \textbf{References} in %\cite{mishchenko2002scattering,antonakakis2014gratings,beutel2021efficient,rahimzadegan2022comprehensive,rahimzadegan2020minimalist,rahimzadegan2021colossal,mun2020describing,koshelev2020dielectric,abujetas2020coupled,evlyukhin2020bianisotropy,beutel2023unified,fradkin2019fourier,de2005full,utyushev2021collective,abujetas2022tailoring,tsang1985,wigner1993matrices,messiah1962clebsch,ewald1921berechnung,kambe1967}
\cite{utyushev2021collective,tsang1985,wigner1993matrices,messiah1962clebsch,ewald1921berechnung,kambe1967,rahimzadegan2021colossal,de2005full}

\bibliographystyle{ScienceAdvances}
\bibliography{References-1}

% Following is a new environment, {scilastnote}, that's defined in the
% preamble and that allows authors to add a reference at the end of the
% list that's not signaled in the text; such references are used in
% *Science* for acknowledgments of funding, help, etc.

% \begin{scilastnote}
% \item We've included in the template file \texttt{scifile.tex} a new
% environment, \texttt{\{scilastnote\}}, that generates a numbered final
% citation without a corresponding signal in the text.  This environment
% can be used to generate a final numbered reference containing
% acknowledgments, sources of funding, and the like, per {\it Science\/}
% style.  Along those lines, we'd like to thank readers of this document
% for their attention, and invite them to address any questions to
% Stewart Wills, at swills@aaas.org.
% \end{scilastnote}

% For your review copy (i.e., the file you initially send in for
% evaluation), you can use the {figure} environment and the
% \includegraphics command to stream your figures into the text, placing
% all figures at the end.  For the final, revised manuscript for
% acceptance and production, however, PostScript or other graphics
% should not be streamed into your compliled file.  Instead, set
% captions as simple paragraphs (with a \noindent tag), setting them
% off from the rest of the text with a \clearpage as shown  below, and
% submit figures as separate files according to the Art Department's
% instructions.

\clearpage

%\begin{appendices}
%\section{Some Appendix}
%The contents...
%\end{appendices}

\end{document}

%\baselineskip24pt
\baselineskip14pt
\maketitle 
%\begin{sciabstract}
%  This document presents a number of hints about how to set up your
%  {\it Science\/} paper in \LaTeX\ .  We provide a template file,
%  \texttt{scifile.tex}, that you can use to set up the \LaTeX\ source
%  for your article.  An example of the style is the special
%  \texttt{\{sciabstract\}} environment used to set up the abstract you
%  see here.
%\end{sciabstract}
%
% In setting up this template for *Science* papers, we've used both
% the \section* command and the \paragraph* command for topical
% divisions.  Which you use will of course depend on the type of paper
% you're writing.  Review Articles tend to have displayed headings, for
% which \section* is more appropriate; Research Articles, when they have
% formal topical divisions at all, tend to signal them with bold text
% that runs into the paragraph, for which \paragraph* is the right
% choice.  Either way, use the asterisk (*) modifier, as shown, to
{\noindent\textbf{\textbf{\large{This PDF includes:}}}\vspace{1mm}\\
\textbf{Sections}\vspace{-1mm}
{\begin{itemize}
\item [-] S1. Field expansion via vector spherical harmonics. \vspace{-2mm}
\item [-] S2. Scattering from 2D arrays via multipolar expansion.\vspace{-2mm}
\item [-] S3. Mie angles definition.\vspace{-2mm}
\item [-] S4. Transformation from Cartesian to Spherical multipoles.\vspace{-2mm}
\item [-] S5. BIC identification for a dipole metasurface.\vspace{-2mm}
\item [-] S6. Computation of the lattice coupling matrix $\bar{\bar{C}}_s$. \vspace{-2mm}
\item [-] S7. Two BICs in one band.
\end{itemize}}
\noindent\textbf{Figures} \vspace{-1mm}
\begin{itemize}
\item [-] Fig.S1. A square lattice of identical particles along with the set-up Cartesian and spherical coordinate systems.\vspace{-2mm}
\item [-] Fig.S2. Model with a Lorentzian dispersion.\vspace{-2mm}
\item [-] Fig.S3. Two BICs on one mode line.\vspace{-2mm}
\end{itemize}
\noindent \textbf{References} \textit{(65-72)}}
\section{Field expansion via vector spherical harmonics}
Assume a particle positioned inside an infinite, non-dispersive, linear, homogeneous, and isotropic medium with an electromagnetic field illuminating it. The total electric field in the spatial domain outside and around the particle at an angular frequency $\omega$ consists of the incident and scattered fields. Each of these fields can be expanded using vector spherical harmonics (VSHs) \textit{(52)} as,
\begin{subequations}\label{fields-vsh}
\begin{equation}
%\begin{split}
\mathbf{E}_{\rm inc}(\mathbf{r}) = 
\sum_{j=1}^{\infty} \sum_{m=-j}^{j} q^{\rm e}_{jm}\mathbf{N}^{(1)}_{jm}(k\mathbf{r}) + q^{\rm m}_{jm}\mathbf{M}^{(1)}_{jm}(k\mathbf{r}),
%\end{split}
\end{equation}
\begin{equation}
%\begin{split}
\mathbf{E}_{\rm sc}(\mathbf{r}) = 
\sum_{j=1}^{\infty} \sum_{m=-j}^{j} a^{\rm e}_{jm}\mathbf{N}^{(3)}_{jm}(k\mathbf{r}) + a^{\rm m}_{jm}\mathbf{M}^{(3)}_{jm}(k\mathbf{r}),
%\end{split}
\end{equation}
\end{subequations}
with $q^v_{jm}$ and $a^v_{jm}$, $j=\{1,2,3...\}$ - positive interger, $m=\{-j,...,j\}$, $v=\{{\rm e,m}\}$, the electric/magnetic incident and scattered field expansion coefficients, respectively, or simply the incident and scattering coefficients. The wavenumber $k$ corresponds to the medium that surrounds the particle, and $r > r_c$, with $r_c$ being the radius of the smallest sphere that circumscribes the particle.  Moreover, the VSH are defined as, 
\begin{subequations}\label{VSW-Hankel}
\begin{gather}
\mathbf{M}^{(l)}_{jm} = \gamma_{jm}\,\nabla\hspace{-0.5mm}\times\hspace{-0.5mm}\left[\hat{\mathbf{r}}z_j^{(l)}\hspace{-0.1mm}(kr)P_j^m( \mathrm {cos} \theta)e^{{\rm i}m\phi}\right],\\
\mathbf{N}^{(l)}_{jm} = \frac{1}{k}\nabla \hspace{-0.5mm}\times\hspace{-0.5mm} \mathbf{M}^{(l)}_{jm},\\
\intertext{with}
\gamma_{jm} =  \sqrt{\frac{(2j+1)}{4\pi j(j+1)}} \sqrt{\frac{(j-m)!}{(j+m)!}},\label{gamma}
\end{gather}
\end{subequations}
with $l=1$ for the incident field and $l=3$ for the scattered one. Additionally, $z^{(1)}_j(x) = j_j(x)$ is the spherical Bessel function of the first kind, while $z^{(3)}_j(x) = h^{(1)}_j(x)$ is the spherical Hankel function of the first kind. Finally, $P^{m}_j(x)$ is the associated Legendre polynomial. 
\section{Scattering from 2D arrays via multipolar expansion}
\begin{figure*}[t]
\begin{center}
\includegraphics[width=0.5\textwidth]{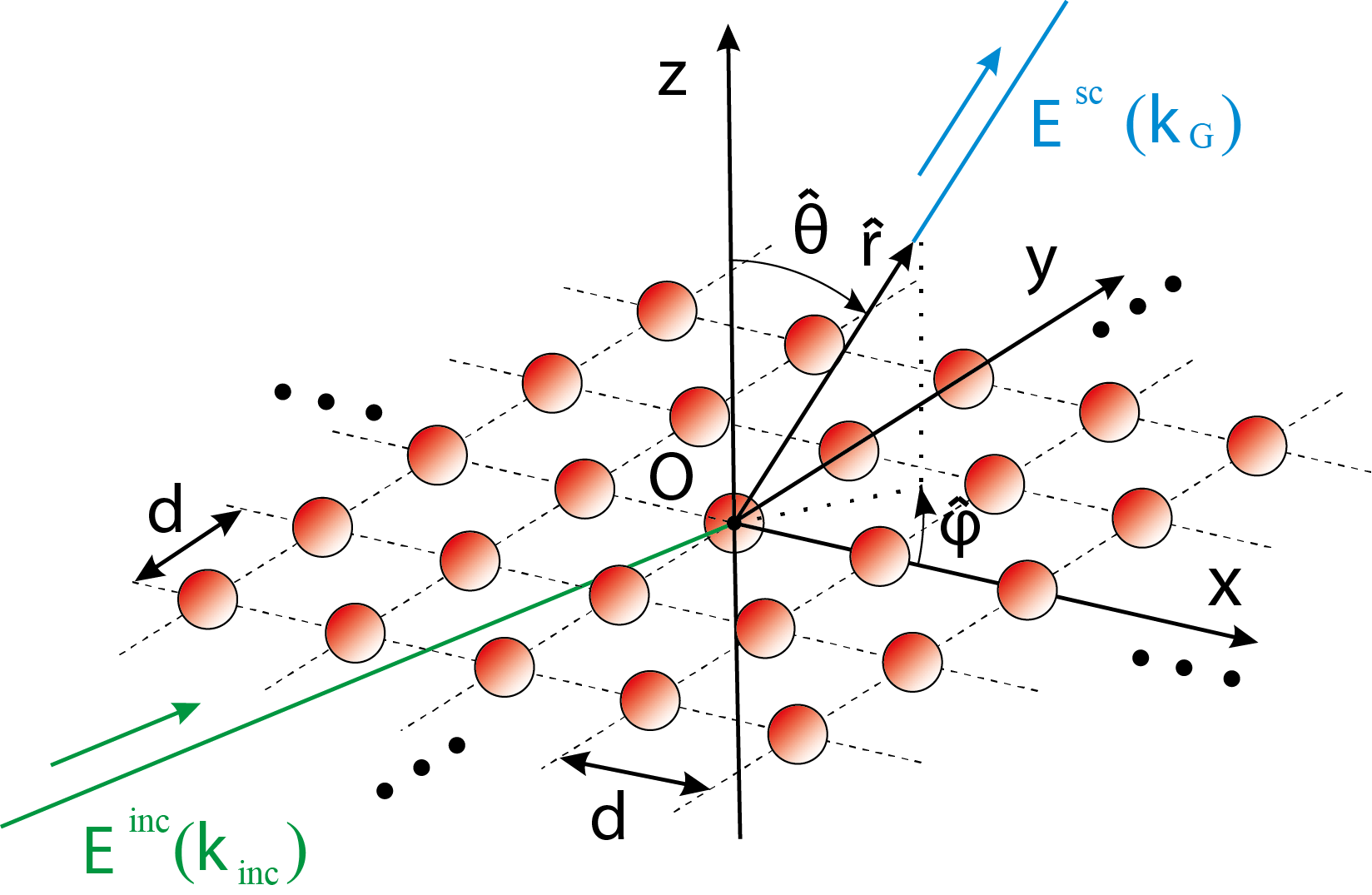}
\caption{A square lattice of identical particles along with the set-up Cartesian and spherical coordinate systems. The 2D array is illuminated by an incident wave, $\textbf{E}^{\rm inc}$, and subsequently emits a scattered wave, $\textbf{E}^{\rm sc}$.} 
\label{fig:2Dlattice}
\end{center}
\end{figure*}
Let us assume an infinite square array composed of arbitrary, identical and absorption-less particles, as displayed in Fig.~\ref{fig:2Dlattice}, placed in a homogeneous material with $\varepsilon_{\rm r}$ and $\mu_{\rm r}$ being the relative permittivity and permeability of the medium, respectively. Let us consider, also, an electromagnetic plane wave impinging onto this 2D array, with an electric field, $\mathbf{E}^{\rm inc} = \mathbf{E}_0\,e^{\mathrm{i}\mathbf{k}^{\rm inc}\cdot\mathbf{r}}$, with a wavevector, $\mathbf{k}^{\rm inc} = {k}^{\rm inc}_x\, \hat{\mathbf{x}} + {k}^{\rm inc}_y\, \hat{\mathbf{y}} + {k}^{\rm inc}_z\, \hat{\mathbf{z}}$, and a magnitude, $E_0=|\mathbf{E}_0|$. 
After employing the VSH formulation and the multipolar expansion method, the scattered fields from each diffraction order, defined by the reciprocal lattice $\textbf{G}$, can be expressed as \textit{(39,40,56)}
\begin{equation}\label{spherical-general-1}
\mathbf{E}^{\rm sc}\hspace{-0.8mm}
=\hspace{-0.8mm}\frac{\mathrm{i}\sqrt{\pi}}{2d^2 k^2} \frac{e^{\mathrm{i}\mathbf{k}_{\mathbf{G}}^{\pm}\cdot\mathbf{r}}}{|\cos{\theta}|} \hspace{-0.5mm}\sum_{j}\hspace{-0.5mm}\frac{\sqrt{2j+1}}{\mathrm{i}^{\,j}} \hspace{-0.5mm} \left[\hspace{-1.2mm}\begin{array}{c}
 \mathbf{W}_j \,  \mathbf{W}^{\prime}_j\\
\mathrm{i} \mathbf{W}^{\prime}_j \, \mathrm{i}\mathbf{W}_j
\end{array}\hspace{-1.2mm} \right]\hspace{-1.2mm} \left[\hspace{-1mm}\begin{array}{c}
\mathbf{a}^{\rm e}_j\\
\mathbf{a}^{\rm m}_j
\end{array} \hspace{-1mm}\right]\hspace{-1mm},
\end{equation}
\noindent where $d$ is the period of the square lattice, $\mathbf{k}^{\pm}_\mathbf{G}$ is the wavevector of the respective mode $\mathbf{G}$,  $\theta$ is the polar angle of the wavevector and  the $\pm$ signs refer to transmission or reflection, respectively. 
The wavevector $\mathbf{k}^{\pm}_\mathbf{G}$ for a square lattice and each of diffraction orders are calculated as (56)
\begin{subequations}\label{modes-wavevectors}
\begin{gather}
\mathbf{k}^{\pm}_\mathbf{G} =  k_{\mathbf{G},x}\,\mathbf{\hat{x}} + k_{\mathbf{G},y}\,\mathbf{\hat{y}} + k^{\pm}_{\mathbf{G},z}\,\hat{\mathbf{z}},\\
%
%\intertext{with}
%
k_{\mathbf{G},x} = k_x^{\rm inc} + \frac{2\pi n_1}{d},\quad k_{\mathbf{G},y} = k_y^{\rm inc} + \frac{2\pi n_2}{d},\\
%\begin{split}
 k_{\mathbf{G},z}^{\pm} \hspace{-0.5mm} =  
\hspace{-0.5mm}\pm\hspace{-0mm} \sqrt{k^2\hspace{-0.5mm} - \hspace{-0.5mm} \left(\hspace{-0.5mm}k_x^{\rm inc} \hspace{-0.5mm} + \hspace{-0.5mm}\frac{2\pi n_1}{d}\hspace{-0.5mm}\right)^2 \hspace{-0.5mm}- \hspace{-0.5mm} \left(k_y^{\rm inc} \hspace{-0.5mm}+\hspace{-0.5mm} \frac{2\pi n_2}{d}\hspace{-0.5mm}\right)^2}
= k\,{\rm cos}\theta,
 %\end{split} 
\end{gather}
\end{subequations}
\noindent where $n_1,n_2\in \mathbb{Z}$ are the diffraction orders.
The respective elements of the vectors $\mathbf{W}_j$ and $\mathbf{W}^{\prime}_j$ are $W_{jm} = \frac{m}{{\rm cos}(\theta)}P^m_j\left({\rm cos} \theta \right)$ and $W'_{jm} = \frac{\partial}{\partial \theta}P^m_j\left({\rm cos} \theta \right)$, with $P^m_j\left(x\right)$ being the associated Legendre polynomial.
The vectors $\mathbf{a}^{\{\rm e,m\}}_j$ in \eqref{spherical-general-1} are the effective electric/magnetic scattering coefficients of the particle inside the lattice, for each multipolar order $j$, and they are calculated via the effective T matrix, i.e. the renormalized T matrix of the constituting particle within the lattice, as presented in (3) of the Main article. 
%\eqref{T-eff-def}.
%
%
\section{Mie angles definition}
Considering an isotropic particle, namely a sphere, made from an absorbing material, its Mie coefficients can be expressed as(42)
\begin{subequations}\label{mie-angles-def-abs}
\begin{gather}
a_j = \frac{1}{1-{\rm i}\,{\rm tan}\theta_{{\rm E}j} + {\rm tan}\theta'_{{\rm E}j}}, \qquad -\frac{\pi}{2}\leq\theta_{{\rm E}j}\leq \frac{\pi}{2},\,\,\,0\leq\theta'_{{\rm E}j}\leq \frac{\pi}{2},\\
b_j = \frac{1}{1-{\rm i}\,{\rm tan}\theta_{{\rm M}j}+ {\rm tan}\theta'_{{\rm M}j}}, \qquad -\frac{\pi}{2}\leq\theta_{{\rm M}j}\leq \frac{\pi}{2},\,\,\,0\leq\theta'_{{\rm E}j}\leq \frac{\pi}{2}.
\end{gather}
\end{subequations}
where $\theta_{{\rm E}j}$ and $\theta_{{\rm M}j}$ are the \textit{detuning Mie angles}, while $\theta'_{{\rm E}j}$ and $\theta'_{{\rm M}j}$ are the \textit{absorption Mie angles}. 

Moreover, if, e.g., the electric dipole Mie coefficient is modelled using a Lorentzian dispersion for a lossless particle, then it can be expressed as a function of the respective Mie angle, or \textit{(42)}, 
\begin{equation}\label{mie-angles-def-dispersion}
%\begin{gather}
a_1 = \frac{\mathrm{i}\gamma^e_r/2}{\left(\omega - \omega_{0e}\right) + \mathrm{i}(\gamma^e_{nr}+\gamma^e_r)/2}, \quad\text{with}\,\,\,
{\rm tan}\theta_{{\rm E}1} = \frac{2\left(\omega - \omega_{0e}\right)}{\gamma^e_r}\,\,\text{and}\,\,{\rm tan}\theta'_{{\rm E}1} = \frac{\gamma^e_{nr}}{\gamma^e_r},
%\end{gather}
\end{equation}
where $\omega_{0e}$ being the resonant frequency and $\gamma^e_r$ and $\gamma^e_{nr}$, the radiative and non-radiative (absorption) losses, respectively, for the specific electric dipole. Other Mie coefficients either magnetic or of higher order can be modelled in a similar fashion with Lorentz dispersion using Mie angles \textit{(42)}. Therefore, the Mie angle formulation can also be useful for the study and design of particles while taking into account a frequency-dependent scattering response.

\begin{figure*}[t]
\begin{center}
\includegraphics[width=0.55\textwidth]{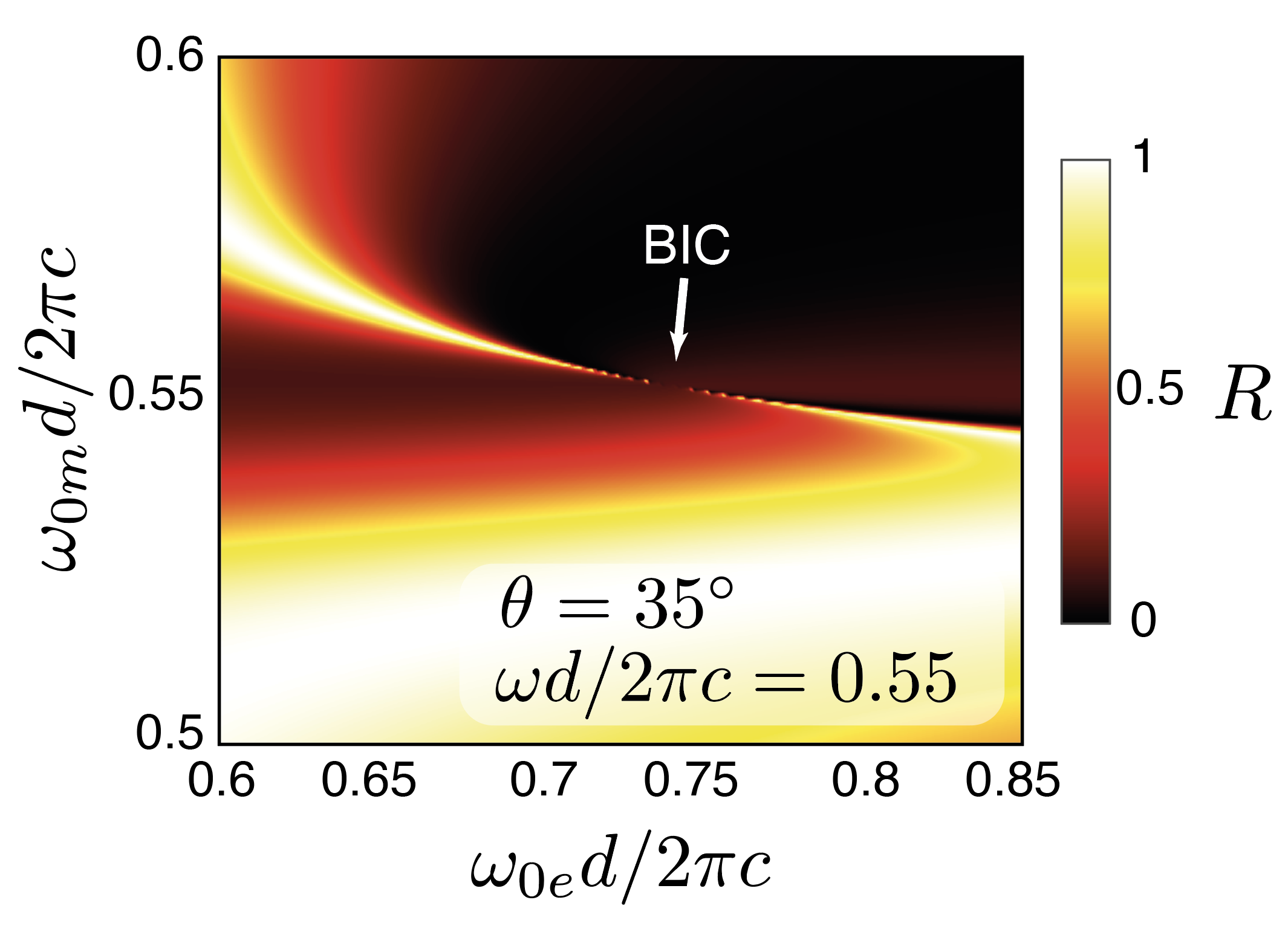}
\caption{Model with a Lorentzian dispersion.  Reflection coefficient magnitude from a cubic 2D array with identical $a_1$ and $b_1$ dipoles exhibiting Lorentz dispersion, as demonstrated in \eqref{mie-angles-def-dispersion}, versus the normalized electric and magnetic resonant frequencies. The scatterers are assumed to be lossless ($\gamma_{nr} = 0$, while the radiative losses are, $\gamma^e_r =1100 c/d $, $\gamma^m_r =360 c/d $, the angle of incidence, $\theta_{\rm inc} = 35^o$, and the normalized frequency, $\omega d/2\pi c = 0.55$.} 
\label{figS1}
\end{center}
\end{figure*}

Let us now examine a BIC example considering Mie coefficients with Lorentz dispersion. If we consider an isotropic particle represented by the Mie coefficients $a_1$ and $b_1$, placed on a cubic lattice with dimension $d$, we can find the BIC point by using \eqref{spherical-general-1}. In particular, assuming no absorption and a non-dispersive radiative losses, one can sweep the resonant frequencies $\omega_{0e}$ and $\omega_{0m}$, for specific operational normalized frequency, $\tilde{\omega}_d = \omega\, d /2 \pi c$,  and angle of incidence, $\theta_d$, as goals, and the resulting reflection coefficient from the corresponding 2D array is depicted in Fig.~\ref{figS1}. A BIC is observed for the resonant frequencies $(\omega_{0e}, \omega_{0m})\cdot d/2\pi c = (0.75,0.55)$, electric and magnetic, respectively, for $\omega d/2\pi c = 0.55$ and $\theta_d = 35^{\rm o}$. Thus, one can, afterward, engineer a particle with the specific parameters using an optimization method to realize this specific BIC behavior with metasurfaces. 

For the case of no losses, the absorption Mie angles turn to $\theta'_{{\rm E}j} = \theta'_{{\rm M}j} = 0$  and \eqref{mie-angles-def-abs}
become the Mie angle expressions for \textit{lossless particles} \textit{(42)}, as applied in the Main article. In particular, when a scatterer is isotropic and without absorption, the respective Mie coefficients can be described in terms of \textit{Mie angles} as 
\begin{subequations}\label{mie-angles-def-S}
\begin{gather}
a_j = \frac{1}{1-{\rm i}\,{\rm tan}\theta_{{\rm E}j}} , \,\, -\frac{\pi}{2}\leq\theta_{{\rm E}j}\leq \frac{\pi}{2},\\
b_j = \frac{1}{1-{\rm i}\,{\rm tan}\theta_{{\rm M}j}} , \,\, -\frac{\pi}{2}\leq\theta_{{\rm M}j}\leq \frac{\pi}{2}.
\end{gather}
\end{subequations}
% = {\rm cos}\theta_{{\rm E}j}\,e^{i\theta_{{\rm E}j},
%= {\rm cos}\theta_{{\rm M}j}\,e^{i\theta_{{\rm M}j}}
The Mie angle formulation enables accessing all possible values of the Mie coefficients for an existing isotropic and without absorption particle, simply by sweeping through all possible angles, $\theta_{{\rm E}j}$ and $\theta_{{\rm M}j}$, as declared in \eqref{mie-angles-def-S}. This versatile formulation can, subsequently, aid the investigation of optical phenomena and design of novel devices \textit{(40,65)}, and is further applied in the Main article.
\section{Transformation from Cartesian to Spherical multipoles} 
Although the scattering response of a single particle and of the subsequent 2D array is provided in this work using the spherical coordinates and the T matrix approach, the proposed analysis is valid and interchangeable with the use of multipoles in cartesian coordinates and the polarizability matrix. 
In particular, one can obtain the T matrix of a particle from the polarizability matrix and vice versa using the appropriate transformation matrices for each multipolar order \textit{(40,58)}. Additionally, modifications of the same  matrices can transform multipoles of the same type and order from a spherical to a cartesian basis and vice versa. For example, in the case of electric dipoles, the transformation between cartesian and spherical coordinates is performed as \textit{(58)},
\begin{subequations}\label{pm-from-ab-1}
\begin{align}
\left[\hspace{-1mm}\begin{array}{c}
p_x\\
p_y\\
p_z
\end{array} \hspace{-1mm}\right] &= \frac{\varepsilon\sqrt{3\pi}}{\mathrm{i}k^3}\left[\hspace{-1mm}\begin{array}{ccc}
1 & 0 & -1 \\
-\mathrm{i} & 0 & -\mathrm{i} \\
0 & \sqrt{2} & 0 
\end{array}\hspace{-1mm}\right]\hspace{-1mm}\left[\hspace{-1mm}\begin{array}{c}
a^{\rm e}_{1-1}\\
a^{\rm e}_{10} \\
a^{\rm e}_{11}
\end{array} \hspace{-1mm}\right],\\
\left[\hspace{-1mm}\begin{array}{c}
m_x\\
m_y\\
m_z
\end{array} \hspace{-1mm}\right] &= -\frac{\sqrt{3\pi}}{\eta k^3}\left[\hspace{-1mm}\begin{array}{ccc}
1 & 0 & -1 \\
-\mathrm{i} & 0 & -\mathrm{i} \\
0 & \sqrt{2} & 0 
\end{array}\hspace{-1mm}\right]\hspace{-1mm}\left[\hspace{-1mm}\begin{array}{c}
a^{\rm m}_{1-1}\\
a^{\rm m}_{10} \\
a^{\rm m}_{11}
\end{array} \hspace{-1mm}\right].
\end{align}
\end{subequations}
Similarly, the transformation above can be performed for quadrupoles, or $j=2$, as \textit{(40,58)}
\begin{subequations}\label{pm-from-ab-2}
\begin{align}
\left[\hspace{-1mm}\begin{array}{c}Q^{\rm e}_{xy}\\ Q^{\rm e}_{yz} \\ \frac{\sqrt{3}}{2}\,Q^{\rm e}_{zz}\\ Q^{\rm e}_{xz} \\\frac{1}{2}\hspace{-1mm}\left(Q^{\rm e}_{xx} - Q^{\rm e}_{yy}\right)\end{array}\hspace{-1mm}\right] &= \frac{\varepsilon\, 6\sqrt{5\pi}}{\mathrm{i}k^4}
\left[\begin{array}{ccccc}
-\mathrm{i} & 0 & 0 & 0 & \mathrm{i} \\
0 & -\mathrm{i} & 0 & -\mathrm{i} & 0 \\
0 & 0 & \sqrt{2} & 0 & 0  \\
0 & 1 & 0 & -1 & 0 \\
1 & 0 & 0 & 0 & 1 
\end{array}\right]\hspace{-1mm}
\left[\hspace{-1mm}\begin{array}{c}
a^{\rm e}_{2-2}\\
a^{\rm e}_{2-1}\\
a^{\rm e}_{20} \\
a^{\rm e}_{21} \\
a^{\rm e}_{22}
\end{array} \hspace{-1mm}\right], \\
\left[\hspace{-1mm}\begin{array}{c}Q^{\rm m}_{xy}\\ Q^{\rm m}_{yz} \\ \frac{\sqrt{3}}{2}\,Q^{\rm m}_{zz}\\ Q^{\rm m}_{xz} \\\frac{1}{2}\hspace{-1mm}\left(Q^{\rm m}_{xx} - Q^{\rm m}_{yy}\right)\end{array}\hspace{-1mm}\right] &= -\frac{6\sqrt{5\pi}}{\eta k^4}
\left[\begin{array}{ccccc}
-\mathrm{i} & 0 & 0 & 0 & \mathrm{i} \\
0 & -\mathrm{i} & 0 & -\mathrm{i} & 0 \\
0 & 0 & \sqrt{2} & 0 & 0  \\
0 & 1 & 0 & -1 & 0 \\
1 & 0 & 0 & 0 & 1 
\end{array}\right]\hspace{-1mm}
\left[\hspace{-1mm}\begin{array}{c}
a^{\rm m}_{2-2}\\
a^{\rm m}_{2-1}\\
a^{\rm m}_{20} \\
a^{\rm m}_{21} \\
a^{\rm m}_{22}
\end{array} \hspace{-1mm}\right].
\end{align}
\end{subequations}
The vectors on the left side of \eqref{pm-from-ab-2} represent one irreducible set of quadrupoles on Cartesian coordinates. The
remaining quadrupoles can be calculated using symmetries, i.e. $Q_{ij} = Q_{ji}$ and $Q_{xx} + Q_{yy} + Q_{zz} = 0$. A relation between octupoles ($j=3$) in Cartesian and spherical coordinates can additionally be found in \textit{(40)}. Extending the transformations for higher orders is, afterward, becoming increasingly difficult and cumbersome, while the intuition provided by the Cartesian coordinates diminishes. Thus, for $j>3$ the use of multipoles represented in spherical coordinates is advised.

When only dipoles are considered,  beginning from \eqref{pm-from-ab-1}, the transformation between the polarizability and T matrices can be derived as \textit{(40,58)}
\begin{subequations}\label{transformation-dipole}     
\begin{equation}
\bar{\bar{\alpha}}^{vv'} = -\frac{\mathrm{i}k^3}{6\pi}\left[\begin{array}{ccc}
1 & 0 & -1 \\
-\mathrm{i} & 0 & -\mathrm{i} \\
0 & \sqrt{2} & 0 
\end{array}\right]\,  \bar{\bar{T}}^{vv'}\, \left[\begin{array}{ccc}
1 & 0 & -1 \\
-\mathrm{i} & 0 & -\mathrm{i} \\
0 & \sqrt{2} & 0 
\end{array}\right]^{-1},\qquad \{v,v'\} = \{{\rm e,m}\}.
\end{equation}
\text{with}
\begin{equation}
\left[\begin{array}{c}
\mathbf{p}  \\
\mathrm{i}\eta\mathbf{m}
\end{array}\right] = \bar{\bar{\alpha}} \left[\begin{array}{c}
\varepsilon\mathbf{E}  \\
\mathrm{i}\eta\mathbf{H}
\end{array}\right]
\end{equation}
\end{subequations}
with the dipolar polarizabilities defined as in \textit{(40)}.
Moreover, if spherical particles are assumed, as it is the case in this work, then, the total polarizability  matrix becomes diagonal, or $\bar{\bar{\alpha}} = 
{\rm diag}\big\{\bar{\bar{\alpha}}^{\rm ee} \,\, \bar{\bar{\alpha}}^{\rm mm}\big\} =
{\rm diag}\big\{\alpha^{\rm ee}_{xx} \,\, \alpha^{\rm ee}_{yy} \,\,\alpha^{\rm ee}_{zz} \,\,\alpha^{\rm mm}_{xx}\,\, \alpha^{\rm mm}_{yy}\,\, \alpha^{\rm mm}_{zz}\big\} = 
{\rm diag}\big\{\alpha^{\rm e} \,\, \alpha^{\rm e} \,\,\alpha^{\rm e} \,\,\alpha^{\rm m}\,\, \alpha^{\rm m}\,\, \alpha^{\rm m}\big\}$. Similarly, the T matrix becomes diagonal, or 
$\bar{\bar{T}} = 
{\rm diag}\big\{\bar{\bar{T}}^{\rm ee} \,\, \bar{\bar{T}}^{\rm mm}\big\} =
{\rm diag}\big\{-a_1 \,\, -a_1 \,\,-a_1 \,\,-b_1\,\, -b_1\,\, -b_1\big\}$, with $a_1$ and $b_1$ being the Mie coefficients, as defined above. Therefore, the polarizabilities of a spherical particle can be directly derived from the Mie coefficients through \eqref{transformation-dipole} as
\begin{equation}\label{sphere-polarizability-mie} 
\alpha^{\rm e} = \frac{\mathrm{i}k^3}{6\pi}\,a_1 \qquad \text{\rm and} \qquad \alpha^{\rm m} = \frac{\mathrm{i}k^3}{6\pi}\,b_1.
\end{equation}
\section{BIC identification for a dipole metasurface}
After having introduced the theoretical tools to describe the electromagnetic response of 2D arrays composed of isotropic particles, in this section, we will describe the procedure of identifying the presence of BICs after setting certain goals.

First, let us theoretically acquire the BIC position using the multipolar expansion technique  \textit{(40,59-61)}. If we again assume a square array composed of identical and isotropic particles (Fig.~\ref{fig:2Dlattice}), its response to an incident field can be described by substituting (1)-(3) from the Main article into \eqref{spherical-general-1}, and, thus, a linear system of equations is formed. Solving the eigenvalue problem leads to the modes of the array, including, in this case, the trapped ones that do not couple with radiation channels, i.e., the BICs. Therefore, if we invert the square matrix of the system to the left side and set the excitation to zero, or $\mathbf{q}^{\rm\{ e,m\}} = \mathbf{0}$, the resulting homogenous system will have a non-trivial solution if the determinant of the matrix is zero. In particular, after some algebra, the BIC condition is reduced to, 
\begin{equation}\label{eig-problem-sph}
\left|\,\bar{\bar{I}} - \bar{\bar{T}}_0\,\bar{\bar{C}}_s\,\right| = 0.
\end{equation}
The equation above is general in nature and can be used for any type of lattice or particle in a homogenous medium \textit{(40)}. Due to its complexity, \eqref{eig-problem-sph} can only be solved numerically in its general form, i.e. for higher order multipoles or more diverse lattices. Nevertherless, for specific reduced cases, versatile analytic solutions can be found, as demonstrated in \textit{(40) }for the case of the coupled electric dipole - magnetic quadrupole on a square lattice and for a TM-polarized plan wave at normal incidence. 

The matrix $\bar{\bar{C}}_s$ is the \textit{lattice coupling matrix} expressed in spherical coordinates, which is a function of the unit cell dimension, the frequency, and the wavevector, and can be calculated via rapidly converging summations using Ewald's method \textit{(39,41,57)}.

In this work, we consider a square lattice decorated by isotropic and lossless particle whose response is expanded only up to dipolar order, or $j=1$. Then, because the elements of the lattice interaction matrix, $\bar{\bar{C}}_s$, can be pre-calculated for a specific incident wavevector, $\mathbf{k}^{\rm inc}$, and a normalized lattice dimension, $d/\lambda$, eventually, \eqref{eig-problem-sph} can be solved with the Mie angles of \eqref{mie-angles-def-S} for $j=1$ as the unknowns. Specifically, for a dipole approximation and an oblique incidence on the lattice, $\bar{\bar{C}}_s$ is simplified to,
\begin{equation}\label{Cs-dip-oblique}
\bar{\bar{C}}_s = \left[\hspace{-0mm}
\arraycolsep=0mm
\begin{array}{cccccc}
C^{\,\rm ee}_{-1-1} & 0 & C^{\,\rm ee}_{-11} & 0 & C^{\,\rm em}_{-10} & 0  \\[0cm]
0 & C^{\,\rm ee}_{00} & 0 & C^{\,\rm em}_{0-1} & 0 & C^{\,\rm em}_{01} \\
C^{\,\rm ee}_{1-1} & 0 & C^{\,\rm ee}_{11} & 0 & C^{\,\rm em}_{10} & 0  \\
0 & C^{\,\rm me}_{-10} & 0 & C^{\,\rm mm}_{-1-1} & 0 & C^{\,\rm mm}_{-11} \\
C^{\,\rm me}_{0-1} & 0 & C^{\,\rm me}_{01} & 0 & C^{\,\rm mm}_{00} & 0  \\
0 & C^{\,\rm me}_{10} & 0 & C^{\,\rm mm}_{1-1} & 0 & C^{\,\rm mm}_{11}
\end{array}\hspace{-0mm}\right] = \left[\hspace{-0mm}
\arraycolsep=0.5mm
\begin{array}{cccccc}
C_{1} & 0 & C_{3} & 0 & C_{5} & 0  \\[0cm]
0 & C_{2} & 0 & C_{5} & 0 & C_{5} \\
C_{3} & 0 & C_{1} & 0 & C_{5} & 0  \\
0 & C_{5} & 0 & C_{1} & 0 & C_{3} \\
C_{5} & 0 & C_{5} & 0 & C_{2} & 0  \\
0 & C_{5} & 0 & C_{3} & 0 & C_{1}
\end{array}\hspace{-0mm}\right].
\end{equation}
Let us now assume an oblique TE (or s-polarized) incident wave. For this specific incidence, a dipole approximation and an isotropic particle only the $m_x$, $p_y$, and $m_z$ dipoles are excited, or, if the spherical coordinates are used, the T matrix turns to $\bar{\bar{T}}_0 = {\rm diag}(-a_1,0,-a_1,-b_1,-b_1,-b1)$. Therefore, \eqref{eig-problem-sph} turns to
\begin{equation}\label{eig-problem-sph-iso}
\begin{split}
\left[(C_1 - C_3)a_1 + 1\right]\cdot&\left[(C_1^2-C_3^2)b_1^2 + 2C_1b_1 + 1\right]\cdot\\
&\left[(C_1 C_2 + C_2 C_3 - 2C_5^2)a_1b_1 + (C_1 + C_3)a_1 + C_2b_1 + 1\right] = 0.
\end{split}
\end{equation}
The first and second terms are associated with collective lattice resonances, which refer to the collective response of the 2D identical particle array \textit{(26,40,66,67)}.  These resonances exhibit no  BIC behavior as they radiate in the environment, and their study is out of the scope of this work. The third term describes the coupling of the $p_y$ and $m_z$ dipoles and, since the accidental BICs rely on the interference between in-plane and out-of-plane multipoles to suppress the far-field radiation, it will provide the condition necessary for the appearance of the BIC mode (see Fig.~2(a)). The same procedure can be repeated for the TM (or p-polarized) incidence, but this time only the $p_x$, $m_y$ and $p_z$ dipoles are excited. Thus, one must insert $\bar{\bar{T}}_0 = {\rm diag}(-a_1,-a_1,-a_1,-b_1,0,-b1)$ in \eqref{eig-problem-sph}.

Following the procedure above, the relation between the dipolar magnetic and electric Mie coefficients is derived for TE or TM incident wave, respectively, from \eqref{eig-problem-sph} as,
\begin{subequations}\label{bic-formulas-dipole-iso}
\begin{align}
b_1 = -\frac{1 + a_1\left(C_{1}+C_{3}\right)}
{C_{2} + a_1\left(C_{1}C_{2}+C_{2}C_{3}-2\,C_{5}^{\,2}\right)},\quad &\text{(TE / s-polarized incidence)} \label{bic-formulas-dipole-iso-a}\\
a_1 = -\frac{1 + b_1\left(C_{1}+C_{3}\right)}
{C_{2} + b_1\left(C_{1}C_{2}+C_{2}C_{3}-2\,C_{5}^{\,2}\right)} \quad &\text{(TM / p-polarized incidence)} \label{bic-formulas-dipole-iso-b}.
\end{align}
\end{subequations}
where $C_{1}$, $C_{2}$, $C_{3}$, and $C_{5}$, are the dipole-dipole interaction coefficients or the elements of the $\bar{\bar{C}}_s$ for $j=1$ for a square lattice \textit{(39,40)}, as demonstrated in \eqref{Cs-dip-oblique}. 
Note that the equations for TE and TM equations are similar with only the coefficients $a_1$ and $b_1$ swapped, as anticipated due to symmetry. One should notice from \eqref{bic-formulas-dipole-iso} that the electric-magnetic lattice coupling coefficient is crucial for the existence of a $(a_1,b_1)$ solution, indicating the importance of multipolar, electromagnetic coupling for realizing a BIC.
%The analytical calculation of \eqref{bic-formulas-dipole-iso} is further elaborated in the \textit{Supplementary Material}. 
 Hence, for the specific aforementioned cases, \eqref{bic-formulas-dipole-iso} gives the exact position of the BIC for given TE and TM incidences, respectively, and they will greatly aid the design and post-processing analysis, herein. In this context, it will become evident why the Mie angle representation of the Mie coefficients is convenient for the identification of the BIC points. For example, for the s-polarized incidence case, by substituting \eqref{mie-angles-def-S} into \eqref{bic-formulas-dipole-iso-a}, and due to the boundaries of $\theta_{\rm E1}$ and $\theta_{\rm M1}$ between $-\pi/2$ and $\pi/2$, the exact solution can be easily obtained via a non-linear equation solver for a given wavelength and lattice dimension. Alternatively, one can sweep $\theta_{\rm E1}$ in \eqref{bic-formulas-dipole-iso-a} between between $-\pi/2$ and $\pi/2$ and keep only the solutions that satisfy the condition $|b_1|=1$ for lossless particles \textit{(42)}, as performed in \textit{(40)}. 
\section{Computation of the lattice coupling matrix $\bar{\bar{C}}_s$}
In this section, we briefly provide the formulas to calculate the elements of a lattice coupling matrix for a 2D multipole array. This discussion is entirely based on our previous work \textit{(39-41)}. The lattice coupling matrix $\bar{\bar{C}}_s$ depends on the dimensions of the 2D array and the relative phase between lattice sites expressed by the components of wave vector $\mathbf{k}_\parallel$ of the incident plane wave that are tangential with respect to the lattice. The vectors $\mathbf{R}=n_1\mathbf{u}_1+n_2\mathbf{u}_2$, $n_1, n_2 \in \mathbb{Z}$ cover all lattice sites with the vectors $\mathbf{u}_1$ and $\mathbf{u}_2$ describing one unit cell of the lattice. We focus now at the particle at the origin. The response of all other sites can be obtained by including the correct phase. The matrix $\bar{\bar{C}}_s$ is calculated after summating the translation coefficients from all other positions on the 2D array to $\mathbf{R} = 0$. 
The radiated fields from a point $\mathbf{R}$ can be reexpanded at the origin using the translation properties of the VSHs \textit{(68)}
\begin{eqnarray} 
\mathbf{M}_{jm}^{(3)}(k{\mathbf{r}}-k{\mathbf{R}})= \sum_{\iota=1}^{\infty}\sum_{\mu=-\iota}^{\iota} 
A_{\iota\mu jm}(-k{\mathbf{R}}) \mathbf{M}_{\iota\mu}^{(1)}(k{\mathbf{r}})
+ B_{\iota\mu jm}(-k{\mathbf{R}}) \mathbf{N}_{\iota\mu}^{(1)}(k{\mathbf{r}}), \\ 
\mathbf{N}_{jm}^{(3)}(k{\mathbf{r}}-k{\mathbf{R}}) = \sum_{\iota=1}^{\infty}\sum_{\mu=-\iota}^{\iota} 
B_{\iota\mu jm}(-k{\mathbf{R}}) \mathbf{M}_{\iota\mu}^{(1)}(k{\mathbf{r}})
+ A_{\iota\mu jm}(-k{\mathbf{R}}) \mathbf{N}_{\iota\mu}^{(1)}(k{\mathbf{r}}),
\end{eqnarray}
\noindent expressed by the translation coefficients
{\small
\begin{subequations} \label{eq:transl}
	\begin{gather}
		\begin{split}
			A_{\iota\mu jm}(kr, \theta, \phi) =
			\frac{\gamma_{jm}}{\gamma_{\iota\mu}}(-1)^m\frac{2\iota+1}{\iota(\iota+1)}\mathrm{i}^{\iota-j}
			\sqrt{\pi\frac{(j+m)!(\iota-\mu)!}{(j-m)!(\iota+\mu)!}}
			\sum_{p} \mathrm{i}^p\sqrt{2p+1} h_p^{(1)}(kr) 
			 Y_{p,m-\mu}(\theta, \phi) \times \\
			\begin{pmatrix}
				j & \iota & p \\
				m & -\mu & -m+\mu
			\end{pmatrix}
			\begin{pmatrix}
				j & \iota & p \\
				0 & 0 & 0
			\end{pmatrix}
			\left[j(j+1)+\iota(\iota+1) - p(p+1)\right],
		\end{split} \\
		\begin{split}
			B_{\iota\mu jm}(kr, \theta, \phi) =
			\frac{\gamma_{jm}}{\gamma_{\iota\mu}}
			(-1)^m\frac{2\iota +1}{\iota(\iota+1)}\mathrm{i}^{\iota-j}
			\sqrt{\pi\frac{(j+m)!(\iota-\mu)!}{(j-m)!(\iota+\mu)!}} 
			\sum_{p} \mathrm{i}^p\sqrt{2p+1} h_p^{(1)}(kr) Y_{p,m-\mu}(\theta, \phi)\times \\
			\begin{pmatrix}
				j & \iota & p \\
				m & -\mu & -m+\mu
			\end{pmatrix} 
			\begin{pmatrix}
				j & \iota & p-1 \\
				0 & 0 & 0
			\end{pmatrix}
			\sqrt{\left[(j+\iota+1)^2-p^2\right]\left[p^2-(j-\iota)^2\right]}.
		\end{split}
	\end{gather}
\end{subequations}}
\noindent with the Wigner 3j-symbols $\begin{pmatrix}
				j_1 & j_2 & j_3 \\
				m_1 & m_2 & m_3
\end{pmatrix}$ \textit{(69-70)}. 
The sum index $p$ takes all integer values for which the Wigner 3j-symbols are non-zero.
The normalization factor $\gamma$ is given in \eqref{gamma}, $h_p^{(1)} (x)$ are the spherical Hankel function of the first kind and
$Y_{p,m-\mu}(\theta,\rho)$ are the spherical harmonics. 
Expressed in matrix form and summing up the response of all other lattice sites to $\mathbf{R} = 0$, $\bar{\bar{C}}_s$ is defined as
\begin{equation} \label{eq:cs:translation}
\bar{\bar{C}}_s = \sum_{\mathbf{R} \neq 0}
	\begin{pmatrix}
		\bar{\bar{A}}(-k\mathbf{R}) & \bar{\bar{B}}(-k\mathbf{R}) \\
		\bar{\bar{B}}(-k\mathbf{R}) & \bar{\bar{A}}(-k\mathbf{R})
	\end{pmatrix}
e^{\rm i \mathbf{k}_\parallel \cdot\mathbf{R}},
\end{equation}
with the rows and columns of $\bar{\bar{A}}$ and $\bar{\bar{B}}$ given by the translation coefficients with $j, \iota \in \mathbb{N}$, $m \in \{-j, -j+1, \dots, j\}$, and $\mu \in \{-\iota, -\iota+1, \dots, \iota\}$.
Now, the crucial step for the numerical evaluation of the matrix coeffiecients is the lattice sum over $\mathbf{R}$. The relevant part of this sum is given by
\begin{equation} \label{eq:dlm}
	D_{jm} = \sum_{\mathbf{R} \neq 0}h_j^{(1)}(kR) Y_{jm}\left(\theta_{-\mathbf{R}}, \phi_{-\mathbf{R}}\right) e^{\rm i \mathbf{k}_\parallel \mathbf{R}}\,.
\end{equation}
The direct evaluation of \eqref{eq:dlm} is generally converging extremely slowely. Therefore, we make use of the \textit{Ewald summation method}  \textit{(71)}. By separating short- and long-range contributions, we can sum them independently in real and Fourier space. This separation leads to two separate quickly converging series. Conventionally writing these parts as $D_{jm} = D_{jm}^{(1)} + D_{jm}^{(2)} + D_{jm}^{(3)}$, where the third summand includes a correction term for the missing origin contribution when converting the long-range part to Fourier space, we get by following \textit{(72)} the expressions
\begin{subequations}\label{eq:dlms}
{\small
	\begin{align}
		D_{jm}^{(1)} &=
		\frac{ \sqrt{(2j+1)(j-m)!(j+m)!}}
		{A k \,\mathrm{i}^{-m}}
		\sum_{\mathbf{G}}
		\left(\frac{|\mathbf{k}_\parallel + \mathbf{G}|}{2k}\right)^j
		\frac{e^{\mathrm{i} m \phi_{\mathbf{k}_\parallel + \mathbf{G}}}}{k_{\mathbf{G},z}^{+}}
		\sum_{\lambda=0}^{\frac{j-|m|}{2}}
		\frac{\left(\frac{k_{\mathbf{G},z}^{+}}{|\mathbf{k}_\parallel +\mathbf{G}|}\right)^{2\lambda} \Gamma\left(\frac{1}{2}-\lambda, -\frac{(k_{\mathbf{G},z}^{+})^2}{4T^2}\right)}
		{\lambda! \left(\frac{j+m}{2} - \lambda\right)!\left(\frac{j-m}{2} - \lambda\right)!},
		\\
		D_{jm}^{(2)} &=
		\frac{-\mathrm{i} (-1)^\frac{j+m}{2} \sqrt{(2j+1)(j-m)!(j+m)!}}{2^{j+1}\pi\frac{j-m}{2}!\frac{j+m}{2}!}
		\sum_{\mathbf{R} \neq 0} e^{\mathrm{i} \mathbf{k}_\parallel\cdot \mathbf{R}+\mathrm{i} m \phi_{-\mathbf{R}}} \frac{1}{k} \left(\frac{2R}{k}\right)^j \int\limits_{T^2}^\infty \, u^{j-\frac{1}{2}} e^{-R^2u+\frac{k^2}{4u}} \mathrm{d} u,\\		
		D_{jm}^{(3)} &= \frac{\delta_{j0}}{4\pi}  \Gamma\left(-\frac{1}{2}\,,\, -\frac{k^2}{4T^2}\right).
	\end{align}}
\end{subequations}
The $D_{jm}^{(1)}$ term involves the reciprocal lattice $\mathbf{G}$ and the wavenumber $k_{\mathbf{G},z}^{+}$ explained in \eqref{modes-wavevectors}, while $\Gamma(\tfrac{1}{2} - \lambda, z)$ is the the upper incomplete Gamma function. 
The parameter $T$ is giving the separation between the real and the Fourier space summation (39) and $A$ is the area of the unit cell defined by $\mathbf{u}_1$ and $\mathbf{u}_2$.
Finally, the integral in $D_{jm}^{(2)}$ is calculated using a recursion relation, while the $\delta_{ij}$ in $D_{jm}^{(3)}$ is the Kronecker delta.
After calculating $D_{jm}$ of \eqref{eq:dlms} with these expressions, one can insert the retrieved value in \eqref{eq:cs:translation} and, thus, obtaining the desired lattice coupling matrix $\bar{\bar{C}}_s$.
 \begin{figure*}[ht!]
\begin{center}
\includegraphics[width=1\textwidth]{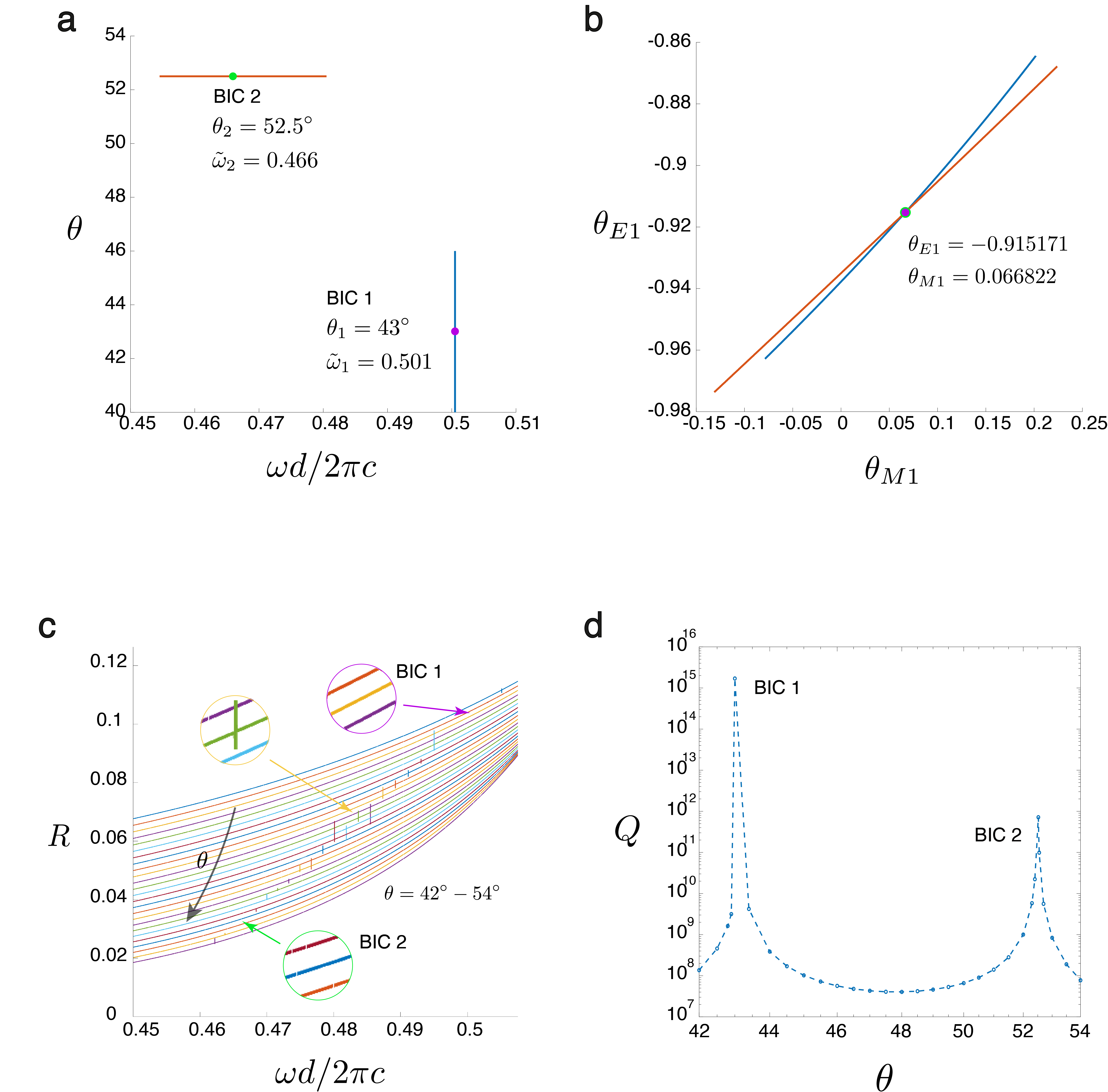}
\caption{Two BICs on one mode line. (a) the positions of the BICs for different angles $\theta$(deg) and frequency $\tilde{\omega}=\omega d/ 2\pi c$ values: first BIC is for $\theta_1= 43\circ$ and $\tilde{\omega}=0.501$, second BIC  is for $\theta_1= 52.5^\circ$ and $\tilde{\omega}=0.466$. (b) the values of the Mie angles for these BICs from panel a. (c)  The reflection R as a function of dimensionless frequency $\omega d/ 2\pi c$ and angle of incidence $\theta = 42^\circ-54^\circ$ for a square lattice with period d = 450 nm. (d) The Q-factor as a function of the angle of incidence $\theta$. } 
\label{fig:pointofcrossline}
\end{center}
\end{figure*}
\section{Two BICs in one band}
During the analysis of BIC robustness in k-space, we found that there is a combination of Mie angles at which two BICs on the same mode line can exist. Figure.~\ref{fig:pointofcrossline}(a) shows the BIC positions for different angles $\theta$ and frequencies $\tilde{\omega}=\omega d/ 2\pi c$. It turns out that there are BICs in different points of k-space that are characterized by the same optical response of a single scatterer, which suggests that these two BICs exist on the same modal line. The first BIC is for $\theta_1= 43^\circ$ and $\tilde{\omega}=0.501$, the second BIC is for $\theta_1= 52.5^\circ$ and $\tilde{\omega}=0.466$. There are two points on the blue and orange line for which the combination of Mie angles is the same (see Fig.~\ref{fig:pointofcrossline}(b)). The Mie angles for this example are $\theta_{E1}  = - 0.915171 , \theta_{M1} = 0.066822$. 
Fig.~\ref{fig:pointofcrossline}(c) shows the reflection $R$ as a function of the normalized frequency $\omega d/ 2\pi c$ and the incident angle $\theta = 42^\circ-54^\circ$ for the square lattice with period $d = 450\, \text{nm}$ and demonstrates that two BICs can exist on one mode line. The green and purple circles mark the areas where the BICs appear. Fig.~\ref{fig:pointofcrossline}(b) shows the dependence of the Q-factor on the angle of incidence $\theta$ and how it changes dramatically for angles $\theta=43^\circ$ and $\theta = 52.5^\circ$.

\clearpage

% \noindent {\bf Fig. 1.} Please do not use figure environments to set
% up your figures in the final (post-peer-review) draft, do not include graphics in your
% source code, and do not cite figures in the text using \LaTeX\
% \verb+\ref+ commands.  Instead, simply refer to the figure numbers in
% the text per {\it Science\/} style, and include the list of captions at
% the end of the document, coded as ordinary paragraphs as shown in the
% \texttt{scifile.tex} template file.  Your actual figure files should
% be submitted separately.

%\begin{appendices}
%\section{Some Appendix}
%The contents...
%\end{appendices}